\documentclass[aps,showpacs,nofootinbib,floats,floatfix,preprintnumbers
, groupedaddress]{revtex4}

\pdfoutput=1
\usepackage{graphicx}
\usepackage{amsmath}
\usepackage{amsfonts}
\usepackage{amssymb}
\usepackage{xcolor, soul}
\usepackage{epstopdf}
\usepackage{comment}
\usepackage{hyperref}
\usepackage{bm}
\usepackage{graphicx}
\usepackage{natbib}
\hypersetup{
  colorlinks   = true,
  citecolor    = red,
  linkcolor    = blue,
  urlcolor     = blue,
}
\usepackage[utf8]{inputenc} 

\def\beq{\begin{equation}}
\def\eeq{\end{equation}}
\def\bea{\begin{eqnarray}}
\def\eea{\end{eqnarray}}

\def\bse{\begin{subequations}}
\def\ese{\end{subequations}}

\def\Tre{T_{\rm re}}

\def\Gev{\mbox{GeV}}

\def\are{a_{\mathrm{re}}}

\def\NI{N_{\mathrm{I}}}

\def\ns{n_{\mathrm{s}}}

\def\HI{H_{_{\mathrm{I}}}}
\def\NI{N_{_{\mathrm{I}}}}

\def\wre{w_{re}}
\def\Nre{N_{\mathrm{re}}}
\def\kre{k_{\mathrm{re}}}

\def\ae{a_{\mathrm{end}}}

\def\l{\left}
\def\r{\right}
\def\ae{a_{\mathrm{e}}}

\def\ere{\eta_{\mathrm{re}}}
\def\Pt{\mathcal{P}_{T}}

\def\nn{\nonumber}
\def\Gk{\mathcal{G}_k}

\def\xe{x_{\mathrm{end}}}

\def\ogw{\Omega_{\mathrm{gw}}}
\def\ogwh{\Omega_{\mathrm{gw}}h^2}

\def\rhogw{\rho_{\mathrm{gw}}}

\def\umin{u_{\rm min}}
\def\umax{u_{\rm max}}

\def\hk{h_{\lambda}}
\def\Ptp{\mathcal{P}^{\lambda}_{_{\rm PRI}}}
\def\Pts{\mathcal{P}^{\lambda}_{_{\rm SEC}}}
\def\Pt{\mathcal{P}^{\lambda}}
\def\ee{\eta_{\rm e}}
\def\Mp{M_{\rm P}}

\def\ere{\eta_{\rm re}}

\def\rhob{\rho_{_{\rm B}}}

\def\vk{\textbf{k}}
\def\vx{\textbf{x}}
\def\vq{\textbf{q}}

\def\vx{\textbf{x}}
\def\gre{g_{\rm re}}
\def\ke{k_{\rm e}}

\def\l{\left}
\def\r{\right}
\def\mH{\mathcal{H}}

\def\Fmunu{F_{\mu\nu}}
\def\Fmunut{F^{\mu\nu}}

\def\mC{\mathcal{C}}
\def\umin{u_{\rm min}}
\def\umax{u_{\rm max}}
\def\mI{\mathcal{I}}

\def\Tre{T_{\rm re}}
\def\gre{g_{\rm re}}

\def\wre{w_{\rm re}}
\def\Teq{T_{\rm eq}}

\def\Gk{\mathcal{G}_k}

\def\kre{k_{\rm re}}

\def\mR{\mathcal{R}}
\def\mP{\mathcal{P}}

\def\mF{\mathcal{F}}
\def\Gev{\text{GeV}}
\def\xe{x_{\rm i}}
\def\ma{m_a}

\def\mS{\mathcal{S}}

\def\ogw{\Omega_{_{\rm GW}}}

\def\rhogw{\rho_{_{\rm GW}}}
\def\Pt{\mathcal{P}^{\lambda}}
\def\Mp{M_{\rm P}}

\def\tx{\tilde{x}}

\def\mPphi{\mathcal{P}_{\Phi}}
\def\mPphip{\mathcal{P}_{\Phi'}}

\def\Ak{A_{\mu}^{\veck,\lambda}}
\def\Akv{A_{\mu}^{(0)\veck,\lambda}}
\def\veck{\mathbf{k}}
\def\vecq{\mathbf{q}}
\def\Akis{A_{i}^{(1)\veck,\lambda}}
\def\Akp{A_{\mu}^{(0)\veck,\lambda}}
\def\Aks{A_{\mu}^{(1)\veck,\lambda}}
\def\Jik{J_i^{\veck,\lambda}}
\def\kp{k_p}
\def\mL{\mathcal{L}}
\def\Tosc{T_{\rm osc}}
\def\eosc{\eta_{\rm osc}}
\def\aosc{a_{\rm osc}}
\def\xosc{x_{\rm osc}}
\def\kosc{k_{\rm osc}}
\def\Hosc{H_{\rm osc}}

\def\eeq{\eta_{\rm eq}}
\def\aeq{a_{\rm eq}}
\def\gs{g_{s,*}}
\def\Pb{\mathcal{P}_{\mathrm{B}}^{\lambda}}

\def\xeq{x_{\rm eq}}
\def\eV{\mathrm{eV}}

\def\Mpc{\mathrm{Mpc}}

\def\mPp{\mathcal{P}_{\Phi}}
\def\mPR{\mathcal{P}_{\mathcal{R}}}
\def\mR{\mathcal{R}}

\def\mC{\mathcal{C}}

\def\mfa{\mathfrak{a}}
\def\fa{f_{a}}
\def\mD{\mathcal{D}}
\def\d{\mathrm{d}}
\def\kpv{k_{*}}
\def\tu{\tilde{u}}

\graphicspath{{./figs/}}
\keywords{
Primordial magnetic field,  Reheating, Primordial Black Holes, Secondary GWs}
\begin{document}

\title{Minimal Magnetogenesis: The Role of Inflationary Perturbations and ALPs,\\ and Its Gravitational Wave Signatures
}
\author{Subhasis Maiti}
\email{E-mail: subhashish@iitg.ac.in}
\affiliation{Department of Physics, Indian Institute of Technology, Guwahati, 
Assam, India}
\author{Rohan Srikanth}
\email{E-mail: rohan.srikanth@uni-potsdam.de}
\affiliation{Institut f{\"u}r Physik und Astronomie, Universit{\"a}t Potsdam, Haus 28, Karl-Liebknecht-Str. 24/25, 14476, Potsdam, Germany}
\author{Debaprasad Maity}
\email{E-mail: debu@iitg.ac.in}
\affiliation{Department of Physics, Indian Institute of Technology, Guwahati, 
Assam, India}

\date{\today}

\begin{abstract}
Any attempt to understand the ubiquitous nature of the magnetic field in the present universe seems to lead us towards its primordial origin.
For large-scale magnetic fields, however, their strength and length scale may not necessarily originate from a singular primordial mechanism, namely inflationary magnetogenesis, which has been a popular consideration in the literature. In this paper, we propose a minimal scenario wherein a large-scale magnetic field is generated from the inflationary perturbation without any non-conformal coupling. Due to their origin in the inflationary scalar spectrum, these primordial fields are inherently weak, with their strength suppressed by the small amplitude of scalar fluctuations. We then consider the coupling between this large-scale weak primordial magnetic field and a light axion of mass $<10^{-28}$ eV, which is assumed to be frozen in a misaligned state until the photon decoupling. After the decoupling, when the universe enters the Dark Ages, the light axion coherently oscillates. By appropriately tuning the axion-photon coupling parameter $\alpha$, we demonstrate that a large-scale magnetic field of sufficient strength can indeed be generated through tachyonic resonance. We further show that the produced magnetic field induces a unique spectrum with multiple peaks of secondary gravitational waves, which the upcoming CMB-S4 can probe through B-mode polarization. The strength can be sufficient enough to violate the PLANCK bound on the tensor-to-scalar ratio $r \lesssim 0.036$. Such a violation leads to a constraint on $\alpha \lesssim 80$. With this limiting value of the coupling, we find that present-day magnetic field strength could be as high as $10^{-10}$ Gauss at $\Mpc$ scale, consistent with observation.              
\end{abstract}
\maketitle

\section{Introduction}
Magnetic fields are ubiquitously present in the Universe at all scales. Despite having dedicated effort over the years, their origin remains an unresolved mystery. Several astrophysical as well as cosmological observations reveal magnetic fields of the order of a few \(\mu \, G\) with coherence lengths of tens to hundreds of kiloparsecs in galaxies and clusters of galaxies \cite{Turner:1987bw, Grasso:2000wj, Giovannini:2003yn, Kronberg:2001st}. Additionally, \(\gamma\)-ray observations originating from distant blazers constrain the magnetic field strength within intergalactic voids to be of the order of \( \sim 10^{-16} \, \mathrm{G}\), with coherence lengths extending up to megaparsec scales \cite{MAGIC:2022piy, Arlen:2012iy, Arlen_2014}. Further, anisotropies in the cosmic microwave background (CMB) provide upper bounds on the strength of primordial magnetic fields (PMFs) at approximately \(1 \, \mathrm{nG}\) on similar scales \cite{Takahashi:2011ac, Arlen:2012iy, paoletti2022constraints, Zucca:2016iur}. These PMFs play a crucial role in various physical processes, such as influencing cosmic ray propagation in galaxies \cite{Aharonian_2011}, altering the dynamics of the primordial plasma in the early universe \cite{Brandenburg1_2020}, and sourcing secondary gravitational waves (SGWs) \cite{Caprini:2014mja, RoperPol:2022iel}.

Despite their significance, a consistent generation mechanism for PMFs remains elusive. Several astrophysical and cosmological models have been proposed. Astrophysical mechanisms typically require seed fields of the order of \(10^{-22} \, \mathrm{G}\), which are then amplified by the galactic dynamo effect \cite{brandenburg:2023}. To explain the origin of such seed fields, cosmological mechanisms have been considered, including phase transitions in the early universe \cite{PhysRevD.58.103505, Kahniashvili:2012uj, Ellis:2019tjf}, magnetic monopoles \cite{Medvedev:2017jdn}, and breaking conformal invariance through couplings to the inflaton field \cite{Kandus:2010nw, PhysRevD.37.2743, Durrer:2013pga, Ferreira:2013sqa, Subramanian:2015lua, Kobayashi:2014sga, Haque:2020bip, Tripathy:2021sfb, Li:2022yqb, Adshead:2016iae}. Other proposals include signatures of multi-inflationary models \cite{Li:2023yqb} and cosmological vector perturbations \cite{Carrilho_2019}. 

 Our work focuses on the inflationary magnetogenesis mechanism, but with a minimally coupled electromagnetic field. We should emphasize that our approach sharply contrasts the conventional inflationary magnetogenesis mechanism with explicit conformal breaking electromagnetism. The electromagnetic field in four dimensions is inherently conformally invariant, and this naturally prohibits the production of classical gauge fields from a quantum vacuum in a conformally flat inflationary background. Therefore, the non-minimal electromagnetic theory is traditionally proposed to have non-trivial production, which explicitly breaks conformal invariance via non-trivial gauge kinetic functions. Those models are mainly divided into two distinct classes depending on the gauge field polarization properties, namely non-helical $f(\eta) FF$ \cite{Martin_2008}, and helical $f(\eta) F \tilde{F}$ \cite{Talebian_2022} type. Apart from the well-known axion ($\mfa$)--photon type coupling function $\mfa\, F \tilde{F}$, any other form of gauge-kinetic functions lacks robust physical motivation due to their ad hoc nature. 
Despite these challenges, a substantial body of work has explored the idea of explicitly breaking of conformal invariance in the electromagnetic sector (see \cite{Tripathy_2022} and references therein). However, this approach continues to face several unresolved issues — notably the strong coupling problem, backreaction effects, and ultraviolet incompleteness — which complicate the construction of consistent magnetogenesis models. Most existing frameworks are primarily aimed at generating magnetic fields of the required strength on Mpc scales directly from inflation. This ongoing effort forms the motivation for our present work.

We divide the generation of the large-scale magnetic field into {\bf two-stage processes}.
The primary goal of such a division would be to formulate a minimal magnetogenesis model in the existing theory of inflation and axion-electrodynamics. As one can understand from the previous discussion, due to the conformal property of the background, the electromagnetic field's growth is shunted in a conformally flat Friedmann–Lemaître–Robertson–Walker (FLRW) universe. However, if the background is conformally non-flat, it is indeed possible to produce gauge field gravitationally without invoking any additional non-minimal coupling. Such scenarios in the context of an anisotropic universe have already been studied in the literature with very limited attention. In particular, for the Bianchi-type anisotropic universe \cite{Parnovsky_2023}, the metric does not remain conformally flat. If such a universe is realized during inflation, a large-scale magnetic field can indeed be produced from the quantum vacuum \cite{Pal:2023vti} with the required strength. However, such models suffer from more fundamental questions related to the realization of an anisotropic universe during inflation itself. Instead, we explore the framework first presented in \cite{Maroto:2000zu}, where conformal inflationary metric perturbations spontaneously break the flatness. In the {\bf first stage} of its evolution, if a minimally coupled gauge field propagates through this perturbed background, a classical electromagnetic field is gravitationally produced from the quantum vacuum at all scales. As the metric perturbation sources the production, the field strength would be naturally Planck suppressed and proportional to the inflationary curvature perturbation strength. However, in the subsequent cosmological evolution, when such a magnetic field encounters a homogeneous oscillating axion field background, our {\bf second stage} of magnetic field generation kicks in. 

The axion is assumed to be a natural extension of the standard model of particle physics, which is invoked to solve the well-known strong CP problem \cite{axions:wilczek}. In addition, it is ubiquitously present in string theory \cite{Svrcek_2006}. It originates from the spontaneous breaking of global U(1) symmetry and hence can be naturally light under a non-perturbative effect. If the axion is initially assumed to be in a misaligned state, it can dynamically evolve to a state of homogeneous oscillating background at a later phase of the universe that can source the resonant production of the large-scale magnetic field. However, for such production, a seed classical gauge field is essential, and the weak magnetic field produced from the inflationary perturbation is set to be the initial condition for such production. The mechanism will be shown to produce a unique helical electromagnetic spectrum. 

At this stage, let us summarize our analysis here. In this scenario, the inflationary inhomogeneity induces mixing between the positive and negative frequency modes of the gauge field, leading to field amplification at superhorizon scales. We then compute the power spectrum of the generated magnetic fields, considering various scalar power spectra, and analyze the relationship between curvature perturbations and magnetic field strength. Once the classical gauge field is produced, it evolves across different cosmic epochs: inflation, reheating, radiation-dominated, and matter-dominated eras. During the matter-dominated phase, once the universe dynamically evolves into an electrically neutral state after recombination, we show that an oscillating axion with appropriate mass can excite the gauge field of cosmologically relevant modes at the Mpc scale, culminating in the present-day magnetic field strength.  
Furthermore, we compute the corresponding gravitational waves (GWs) arising from the late-time excitation of the gauge field due to axion oscillations. Our findings indicate that these GWs can leave a non-trivial imprint on the GW spectrum at CMB scales, which could be detectable by future CMB experiments such as CMB-S4~\cite{CMB-S4:2020lpa}.  
Interestingly, this mechanism inherently produces a helical magnetic field. Since helical magnetic fields act as a source of GWs, the resulting GWs are also expected to be helical. Additionally, we find that the magnetic field exhibits a spectral peak whose position depends on the axion mass, leading to a distinctive feature in the GW spectrum. Such a peak could imprint a unique signature on the CMB spectrum, offering a potential observational probe of this mechanism. Such observable imprints in the CMB we leave for our future studies.

The paper is structured as follows. In Section~\ref{sec_II}, we present the general formalism of gauge field production from the inflationary curvature power spectrum. Additionally, we further show how different types of inflationary power spectra, such as primordial black hole (PBH)-type curvature power spectrum, can impact the magnetic field generation. The PBH power spectrum is usually characterized by a peak frequency. We analyze how the spectral behavior of the magnetic field is modified near the peak and discuss how the peak position influences the superhorizon magnetic fields. Additionally, we examine the impact of different reheating scenarios on the present-day magnetic field.  
In Section~\ref{sec_III}, we explore how, after the recombination epoch, the presence of ultralight axions (ULAs) can lead to a resonant amplification of the seed gauge field generated by the inflationary curvature power spectrum to an observationally favored value at $\Mpc$ scale. Section~\ref{sec_IV} discusses the potential detection of this mechanism through the secondary gravitational waves generated from the resonantly produced magnetic field in the homogeneous oscillating axion background. Finally, in Section~\ref{sec_V}, we summarize our key findings.

\section{First stage: {\rm Magnetogensis from curvature power spectrum }}\label{sec_II}

\subsection{Generalities} In the standard scenario, the conformal flatness of the background and conformal invariance of the electromagnetic field conspire in preventing the production of the electromagnetic field during inflation. However, in the presence of curvature perturbations, the conformal flatness of the background is dynamically broken, leading to the generation of primordial magnetic fields. In this section, we explore the evolution of the gauge field in such a scenario following the reference \cite{Maroto:2000zu}, where curvature-induced modifications to the electromagnetic action allow for efficient field amplification. So we begin with the following equation
\begin{align}
      \mathcal{S} = \int  d^4x \ \sqrt{-g} \left[ \frac{M^2_{pl}}{2} R +\frac{1}{2} \partial^\mu \varphi \partial_\mu \varphi - V(\varphi) - \frac{1}{4} F_{\mu \nu} F^{\mu \nu} \right].
\end{align}
Now, finding the equation of motion for the gauge field, we arrive at,
\begin{equation}
\partial_\nu \left( \frac{\delta \mathcal{L}_{EM}}{\partial \partial_\mu A_\nu } \right)= \partial_\nu \left( \frac{\partial \sqrt{-g} F_{\mu\nu} F^{\mu\nu}}{\partial \partial_\mu A_\nu} \right) = \frac{\partial}{\partial x^\mu} [\sqrt{-g} g^{\alpha \mu}g^{\beta \nu} F_{\alpha \beta}] = 0,
\end{equation}
To begin with, we consider the following perturbed flat FLRW metric, 
\begin{align}
    ds^2=a^2(\eta)\l[-(1+2\Psi)d\eta^2 +(1+2\Phi)\delta_{ij}dx^idx^j \r]
\end{align}
In this background, for $\nu=0$, we find,
\begin{align}
    \frac{\partial}{\partial x^i}\l[ (1-2\Phi)(\partial_iA_0-\partial_0 A_i) \r]=0 \label{eq_nu_0},
\end{align}
and for $\nu=i$ we get,
\begin{align}
    \frac{\partial}{\partial\eta}\l[ (1-2\Phi)(\partial_i A_0-\partial_0 A_i)\r] +\frac{\partial}{\partial x^j}\l[ (1+2\Phi)(\partial_jA_i-\partial_i A_j )\r] =0 \label{eq_nu_i}.
\end{align}
In addition, we have considered the Coulomb gauge condition $\vec{\nabla}\cdot \vec{A}=0$. In the limit of vanishing anisotropic stress, we further assume $\Psi = -\Phi$. From the above expression, it is clear that in the presence of $\Phi$, the gauge field should amplify itself from the vacuum sourced by the metric perturbation. If we assume that in the asymptotic past, all the modes were well inside the horizon and choose the well-known Bunch-Davis (BD) vacuum, we can write
\begin{align}
    \Ak(x) \xrightarrow[\eta \to -\infty]{} \Akv(x)= \frac{1}{\sqrt{2k}}\epsilon_\mu(\veck,\lambda)e^{i({\vec k}\cdot {\bf x}-k\eta)}\label{eq_A_vac},
\end{align}
where $\epsilon_\mu(\veck,\lambda)$ is the polarization vector for gauge field. Now, after inflation starts, an inhomogeneous background is formed, and in the presence of a non-vanishing metric perturbation $\Phi$, the gauge field is excited from the BD vacuum. Therefore, in the asymptotic future, this solution will behave as a linear superposition of positive and negative frequency modes with different momentum and polarization. We can express this gauge field as
\begin{align}
    \Ak(x)\xrightarrow[\eta \to \infty]{} \sum_{\lambda'}\sum_{q} \l( \alpha_{\veck}^{\lambda\lambda'}(q,\eta) \frac{\epsilon_{\vq}^{\lambda}}{\sqrt{2q}} e^{i(\vq\cdot\vx-q\eta)} + \beta_{\veck}^{\lambda\lambda'}(\vecq,\eta) \frac{\epsilon^{\lambda *}_{\vecq}}{\sqrt{2q}}e^{-i(\vecq\cdot\vx -q\eta)} \r).\label{eq:Ak_beta}
\end{align}
Here, `$\alpha_{\vk}^{\lambda\lambda'}$' and `$\beta_{\vk}^{\lambda\lambda'}$' can be taken to be as the well-known Bogolyubov coefficients, and we can easily determine the expression order by order in terms of metric perturbation. In order to compute $\beta_{\vk}^{\lambda\lambda'}(q,\eta)$, we expand the gauge field as \(\Ak(\vecq,\eta)=\Akp(\vecq,\eta)+\Aks(\vecq,\eta)+\cdots\). Here, $\Akp$ is the solution of the gauge field in the absence of perturbation. Utilizing this with Eqs. (\ref{eq_nu_0})(\ref{eq_nu_i}) (\ref{eq_A_vac}), we get the equation of motion of the gauge field in Fourier space as
\begin{align}
    \frac{d^2}{d\eta^2}\Akis(\vecq,\eta)+q^2\Akis(\vecq,\eta)=\Jik(\vecq,\eta)\label{eq_Ak_s},
\end{align}
where the source term $\Jik(\vecq,\eta)$ takes the following form \cite{Maroto:2000zu},
\begin{align}
    \Jik(\vecq,\eta)=-\sqrt{2q}\l[ \l( i\Phi'(\veck+\vecq,\eta)+\frac{q^2-\veck\cdot\vecq}{q}\Phi(\veck+\vecq,\eta)\r)\epsilon_{i\veck}^{\lambda} e^{-iq\eta} +( \epsilon_{\veck}^{\lambda}\cdot\vecq) \Phi(\veck+\vecq,\eta) \frac{k_i}{k} e^{-iq\eta}\r.\nn\\
   \l. -i\frac{\epsilon_{\veck}^{\lambda}\cdot\vecq}{q^2}\frac{\partial}{\partial \eta}\l[ \Phi(\veck+\vecq,\eta)e^{-ik\eta} \r]\vecq_i\r] .\label{eq_Jik}
\end{align}
Detailed calculations are given in the Appendix \ref{appendix-A}. Now, we solve the above equation using the Green's function method, and we can write the solution of the gauge field as
\begin{align}
   A_i^{\veck,\lambda}(\vecq,\eta)=\frac{\epsilon_{i\veck}^{\lambda}}{\sqrt{2k}}\delta(\veck-\vecq)e^{-ik\eta}+\frac{1}{2}\int_{\eta_i}^{\eta_f}\Jik(\vecq,\eta)\mathcal{G}_q(\eta_f,\eta')d\eta'\label{eq_sol_Ak} ,
\end{align}
where $\mathcal{G}_q(\eta_f,\eta')$ is the Greens function of Eq.(\ref{eq_Ak_s}). The first term of Eq.(\ref{eq_sol_Ak}) is due to the homogeneous part of Eq.(\ref{eq_Ak_s}) and can be assumed as quantum fluctuations as long as they are inside the horizon during inflation. Now the Green's function of the corresponding Eq.(\ref{eq_Ak_s}) is $\mathcal{G}_q(\eta_f,\eta')=\Theta(\eta_f-\eta')\sin(q(\eta_f-\eta'))/q$. Utilizing this in Eq.(\ref{eq_sol_Ak}) and compared with Eq.(\ref{eq:Ak_beta}) it is straight forward to obtain the Bogoliubov coefficient $\beta^{\lambda\lambda'}_{\veck}(\vecq,\eta)$ as,
\begin{align}
    \beta_{\veck}^{\lambda\lambda'}(\vecq,\eta) =-\sum_i \frac{i}{\sqrt{2q}}\int_{\eta_i}^{\eta_f}\epsilon^{\lambda *}_{i\vecq} \Jik(\veck+\vecq,\eta')e^{-iq\eta'}d\eta' .
\end{align}
Where $\eta_i$ and $\eta_f$ are the initial time when the source is active and the final time, respectively. Once we know the behavior of the metric fluctuation, then from the above integral we can easily compute the total number of photons created with the comoving wave number $q$, and the expression is given by $N_q=\sum_{\lambda\lambda'}\sum_{k}\l|\beta_{\veck}^{\lambda\lambda'}(\vecq,\eta) \r|^2$. In the continuum limit, we can write the total number of photons in terms of the power spectrum as (see detailed calculation in appendix \ref{appendix-A})
\begin{align}
    N_k=\sum_{\lambda\lambda'}\int \frac{d^3q}{(2\pi)^{3}}\Phi_0^2(|\veck+\vecq|) {\cal D}(k,q), 
    \end{align}
    where we introduce
    \begin{align}
 {\cal D}(k,q) =  \l|\int d\eta' \l\{ \l( i \mC^{'}(|\vk+\vq|\eta') + \frac{q^2-\veck\cdot\vecq}{q}\mC(|\vk+\vq|\eta') \r)(\epsilon^{\lambda}_{\veck}\cdot\epsilon^{\lambda'}_{\vecq}) +
    \frac{(\epsilon^{\lambda}_{\veck}\cdot\vecq)(\epsilon^{\lambda'}_{\vecq}\cdot\veck)}{q^2}\mC(|\vk+\vq|\eta')  \r\}\r|^2   .
\end{align}
Here, $\mC(k\eta)$ can be identified as the transfer function defined as $\Phi(q,\eta)=\Phi_0(q)\mC(q\eta)$, and ${\cal C}'(q\eta) = d{\cal C}/{d\eta}$. Typically the scalar power spectrum is expressed as \(\mPphi(\veck+\vecq)=(|\veck+\vecq|)^3 \Phi_0^2(|\veck+\vecq|)/2\pi^2\). As we have seen, the last term of Eq.(\ref{eq_Jik}) does not contribute to the number spectrum due to the transversality condition of the polarization vectors.

\paragraph{Defining the Present-day magnetic field Strength:\\}
Once we compute the total photon number density for the comoving wave number `$k$', then we can immediately compute the magnetic energy density in terms of the scalar power spectrum stored in the mode `$k$' at any instant of time as
\begin{align} \label{magpower}
    \mathcal{P}_{\rm B}(k,\eta)=\frac{\partial\rho_{\rm B}(\eta)}{\partial\ln(k)}=\l(\frac{k}{a(\eta)}\r)^4 N_k(\eta) = \l(\frac{k}{a(\eta)}\r)^4 \sum_{\lambda\lambda'}\int \frac{d^3q}{(2\pi)^{3}} \l(\frac{q}{k}\r)\frac{2\pi^2 \mPphi(\veck+\vecq)}{(|\veck+\vecq|)^3} {\cal D}(k,q) 
\end{align}
 here `$a(\eta)$' is the scale factor defined at `$\eta$'.  After the production of the gauge field, it evolves adiabatically, and due to the expansion of our universe, the magnetic energy density dilutes as $a^{-4}$. If there is no additional source for the magnetic field during the subsequent evolution, we can define the present-day magnetic field as
\begin{align}
    B_0(k)=\l(\frac{k}{a(\eta_0)}\r)^2\sqrt{N_k},
\end{align}   
We set $a(\eta_0)=1$ to define the present-day scale factor.

\paragraph{Defining the Inflationary and Reheating Parameters:} 
We shall now establish a connection between inflationary parameters and reheating dynamics, which are typically characterized by two key parameters: the average equation of state (EoS), denoted as $\wre$, and the reheating temperature $\Tre$, which is directly related to the duration of the reheating era. 

In a simple scenario, inflationary dynamics can be parameterized by several quantities. However, in our context, two crucial aspects are required: (i) the inflationary energy scale $\HI$ and (ii) the duration of the inflationary period, expressed in terms of the comoving wavenumbers, $\ke=k_*\mathrm{Exp}[\NI]$,
where  $k_* \simeq 0.05~\text{Mpc}^{-1}$ is the CMB pivot scale, $\NI$ represents the total number of e-folds during inflation, and $\ke$ is the wavenumber exiting the horizon at the end of inflation. For our entire subsequent calculation, we assume an inflationary energy scale of $\HI\simeq 10^{-5}~\Mp$ consistent with PLANCK observation, where  $\Mp = 1/\sqrt{8\pi G}$ is the reduced \textit{Planck} mass. 

The duration of inflation, quantified through $\ke$, can be expressed in terms of the reheating parameters $\Tre$ and $\wre$, following the standard Kamionkowski-type model~\cite{PhysRevLett.113.041302}. Assuming that the reheating dynamics are governed by an average equation of state $\wre$, the evolution of the inflaton, denoted by say $\phi$, energy density during this phase is given by $\rho_{\phi} = \rho_{\phi}^{\rm end} \left({a_e}/{a}\right)^{3(1+w_{\wre})}$,
where $\rho_{\phi}^{\rm end} = 3\HI^2\Mp^2 $ is the total inflaton energy density at the end of inflation.

The reheating temperature \( T_{\text{re}} \) is defined as the temperature of the Universe when reheating ends. Reheating is assumed to end at conformal time \( \eta = \eta_{\text{re}} \), at which point the inflaton energy density equals the radiation energy density, i.e., \( \rho_\phi(\eta_{\text{re}}) = \rho_{\text{ra}}(\eta_{\text{re}}) \)~\cite{PhysRevLett.113.041302}. Using this condition, the reheating temperature \( T_{\text{re}} \) can be expressed as~\cite{PhysRevLett.113.041302, Maiti:2024nhv}
\begin{align}
\Tre = \left(\frac{90\HI^2\Mp^2}{\pi^2 \gre}\right)^{1/4} \exp\left[-\frac{3\Nre}{4}(1+\wre)\right],\label{eq:tre}
\end{align}
where \( g_{\text{re}} \) represents the number of relativistic degrees of freedom at the end of reheating. 

Alternatively, the duration of the reheating period can be rewritten as~\cite{PhysRevLett.113.041302, Chakraborty:2024rgl}
\begin{align}
\Nre = \frac{1}{3(1+\wre)}\ln{\left(\frac{90\HI^2\Mp^2}{\pi^2 \gre\Tre^4}\right)}.\label{eq:nre}
\end{align}

Assuming negligible entropy production after reheating, leading to the conservation of comoving entropy density (\( a^3(\eta \geq \ere)s = \text{const} \)), we can establish a relationship between the smallest mode re-entering the horizon at the end of reheating and the reheating temperature as~\cite{Chakraborty:2024rgl}
\begin{align}\label{eq:kre}
    \l(\kre/a_0\r)\simeq 1.82\times 10^5~\l(\frac{\Tre}{10^{-2}~\text{GeV}}\r)~\text{Mpc}^{-1} .
\end{align}
By utilizing Eq.~(\ref{eq:nre}), the largest mode that exited the horizon at the end of inflation can be expressed as~\cite{Chakraborty:2024rgl}
\begin{align}
     \l(\ke/a_0\r)=\left(\frac{43 }{11\gre}\right)^{1/3}\left(\frac{\pi^2\gre}{90}\right)^{\alpha_1}\frac{\HI^{1-2\alpha_1}\Tre^{4\alpha_1-1}T_0}{\Mp^{2\alpha_1}},\label{eq:ke}
\end{align}
where \( \alpha_1 = 1/3(1+w_{\text{re}}) \), \( T_0 = 2.725 \) K is the present-day CMB temperature, and \( a_0 = 1 \) is the present-day scale factor.

\subsection{Generation of the magnetic field during the inflation era}
In this subsection, we describe the evolution of the magnetic field power spectrum and number density during inflation. Considering a simple slow-roll type inflation with minimally coupled potential, one can write the solution for $\Phi$ following \cite{mukhbranden},
\begin{equation}
    \Phi^{''} + 2\left(\mathcal{H} - \frac{\varphi^{''}}{\varphi}\right)\Phi^{'} - \nabla^2 \Phi +  2\left(\mathcal{H}^{'} - \mathcal{H}\frac{\varphi^{''}}{\varphi}\right) \Phi = 0,
\end{equation}
where $\mathcal{H}$ is the Hubble parameter with respect to conformal time $\eta$, and $'$ indicates the derivative with respect to $\eta$. It is more convenient to revert to physical time 't'. 
In the sub-horizon scale, the solution is oscillatory in time. In the superhorizon scale, considering the background equation of motion, slow roll conditions during inflation, and rewriting in terms of physical time, one gets
\begin{equation}
    \Phi \equiv - A \frac{\Dot{H}}{H^2} \sim \frac{A}{16\pi G} \left(\frac{V_{,\phi}}{V} \right)^2.
\end{equation}
Or, in terms of the curvature perturbation $\mathcal{R}$,
\begin{align}
    \Phi \sim \frac{\sqrt{\epsilon}}{2\sqrt{k^3}} \frac{H}{M_p} \sim \epsilon \mathcal{R},
\end{align}
where $\epsilon$ is the slow-roll parameter during inflation which suppresses $\Phi$ during inflation. Further, in terms of the equation of state, on superhorizon scales \cite{mukhbranden}, 
\begin{equation}
    \mathcal{R} = \frac{\frac{2}{3} H^{-1} \dot{\Phi}+\Phi}{(1+w)} + \Phi
\end{equation}
Despite the suppression from $\epsilon$, one can still compute the total number of photons generated due to inflation purely arising from the metric fluctuations $ \sim \Phi$ contribution as
\begin{equation}
    N_k=\sum_{\lambda\lambda'}\int \frac{d^3q} {(2\pi)^{3}} \epsilon^2 \Phi_0^2(|\veck+\vecq|) {\cal D}(k,q), 
\end{equation}
where ${\cal D}(k,q)$ is a function of the momenta and the transfer function as previously mentioned. $\Phi$ oscillates and decays on sub-horizon scales, and the contribution is negligible. on super-horizon scales the behavior depends on that of $\epsilon$, as $\mathcal{R}$ is roughly constant. For the simple quadratic potential, nearly scale-invariant power spectrum, we begin with splitting the above integral into two limit $\int_{\kpv}^{\ke}\d q=\int_{\kpv}^k\d q+\int_k^{\ke}\d q$, where the limits indicate the peak wavenumber and the wavenumber corresponding to the CMB, and we can write $N_k$ as 
\begin{align}
    N_k &\simeq\int_{\kpv}^k\frac{\d q}{q}\frac{q}{k}\frac{q^3}{2\pi^2}\epsilon^2\mR^2(k)\times\mathcal{O}(1)+\int_{k}^{\ke}\frac{\d q}{q}\frac{q}{k}\frac{q^3}{2\pi^2}\epsilon^2\mR^2(k)\times\mathcal{O}(1),
\end{align}
and the magnetic power spectrum takes the form,
\begin{align}
    \mP_{\rm B}(k,\ee)\simeq \frac{k^3\kpv}{a^4(\ee)}\frac{\epsilon^2A_s}{\ns}\l\{ \l(\frac{\ke}{\kpv}\r)^{\ns}\r\}\label{eq:mPbk_i2},
\end{align}

We can further consider an ultra-slow-roll type inflationary scenario, where the enhanced curvature power spectrum is generated due to the ultra-slow-roll phase, and the comoving curvature power spectrum is parameterized as
\begin{align}
      \mathcal{P_R}(k) = A_s \left(\frac{k}{k_*}\right)^{n_s-1} + A_0~\text{Exp}\left[-\frac{(k-k_p)^2}{\delta\times k_p^2}\right],\label{eq:def_PR1}
\end{align}
Here $A_s\simeq 2.1\times 10^{-9}$ is the amplitude of the scalar curvature power spectrum observed by Planck, and $\ns$ is the scalar spectral index~\cite{Planck:2018vyg}. Here, $A_0$ is the amplitude of the curvature power spectrum at peak wavenumber, and $\delta$ is the dimensionless parameter that sets the spectral shape of the curvature near the peak. $\kp$ is the position of the peak of the curvature power spectrum. In this case, we get two contributions to the number density - from the usual nearly-scale invariant term in the first part in \ref{eq:def_PR1}, and the PBH-type enhancement in the second term. Hence, the total number density is of the form,
\begin{align}
    N_k\simeq \frac{\epsilon^2A_s}{\ns}\l(\frac{\kpv}{k}\r)\l\{ \l(\frac{\ke}{\kpv}\r)^{\ns}+\frac{A_0\ns\sqrt{\pi\delta}}{A_s}\l(\frac{\kp}{\kpv}\r)\l[ 1+\text{Erf}\l(\frac{\kp-k}{\kp\sqrt{\delta}}\r)\r]\r\}.
\end{align}
As we see, the total number produced for co-moving wavenumber $k$ scales as $N_k\propto k^{-1}$. The energy density of the produced magnetic field per logarithmic wavenumber is defined as $\rhob(k,\eta)=(k/a(\eta))^4N_k(\eta)$. Or we can define the comoving magnetic energy density as $\tilde{\rhob}(k,\eta)=k^4N_k(\eta)$. Now, at the end of inflation, the magnetic energy store per logarithmic $k$ is
\begin{align}
    \mP_{\rm B}(k,\ee)\simeq \frac{k^3\kpv}{a^4(\ee)}\frac{\epsilon^2A_s}{\ns}\l\{ \l(\frac{\ke}{\kpv}\r)^{\ns}+\frac{A_0\ns\sqrt{\pi\delta}}{A_s}\l(\frac{\kp}{\kpv}\r)\l[ 1+\text{Erf}\l(\frac{\kp-k}{\kp\sqrt{\delta}}\r)\r]\r\}\label{eq:mPbk_i}.
\end{align}

\subsection{Generation of magnetic field for post-inflationary evolution}

As shown above, the magnetic power spectrum at any given time is proportional to the inflationary scalar power spectrum. After the end of inflation, this scalar power spectrum evolves into the curvature power spectrum, which can, in principle, influence the post-inflationary evolution of the gauge field.

In the high-conductivity limit, however, the production of gauge fields by any source term becomes highly suppressed. During reheating, the generation of charged particles—and hence the development of conductivity- depends on the specific model describing how the inflaton decays into Standard Model particles. In simple perturbative reheating scenarios, the energy density remains largely dominated by the inflaton field for most of the duration, making the behavior of conductivity during this phase uncertain.

If conductivity develops rapidly after inflation and is assumed to affect all comoving modes k uniformly, then any further gauge field production would be strongly suppressed. Given the lack of precise knowledge about the evolution of conductivity during reheating, we assume a small but nonzero electric conductivity during this phase. Full conductivity is considered to be established only after the inflaton has completely decayed and the universe becomes thermalized, entering a radiation-dominated epoch.

In this section, we consider different possible inflationary spectra consistent with CMB and explore their impact on the present-day magnetic field strength. Before we do that, we provide a quick summary of the post-inflation evolution of the scalar perturbation $\Phi$. For this, let us assume that after inflation, the universe is filled with a fluid of equation of state $w$, and the energy density of the fluid evolves as ${H}^2 \propto \rho \propto a^{-3(1+w)}$. For example, the reheating phase corresponds to $w =\wre$, radiation domination corresponds to $w=1/3$.
In the linear order in perturbation theory, the equation of motion (EoM) for $\Phi$ after inflation in Fourier space is given by \cite{Domenech:2020kqm}
\begin{align}
      \Phi''+\frac{4+2\beta}{1+\beta}\mH \Phi'+k^2\Phi=0,
\end{align}
Here $\beta=(1-3w)/(1+3w)$ and the solution of the above equation is well known and takes the form\cite{Domenech:2020kqm},
\begin{align}
    \Phi(k,\eta)=\Phi_0(k)2^{\beta+3/2} \Gamma[\beta +5/2](k\eta)^{-\beta-3/2} J_{\beta+3/2}(k\eta)=\mC(k\eta)\Phi_0(k).\label{eq:phi_sol}
\end{align}
Where the primordial part $\Phi_0(k)$ is set by inflationary dynamics, and is related to the curvature perturbation on co-moving constant energy slices, and $\mC(k\eta)$ is the transfer function associated with a given post-inflationary phase with an equation of state $w$.

We can now write the primordial scalar power spectrum $\mPp(k)$ in terms of comoving curvature power spectrum $\mPR(k)$ as
\begin{align}\label{eq:P_phi_post_i}
    \mPp(k,\eta)=\l( \frac{2+\beta}{3+2\beta}\r)^2\l|\mC(k\eta)\r|^2 \mathcal{P_R}(k) .
\end{align}
As long as a particular mode is outside the horizon during the phase under consideration, namely in the $k\eta<<1$ limit (super-horizon limit), the transfer function freezes to unity $\l|\mC(k\eta)\r|^2 \simeq 1$, and we have,
\begin{align}
    \mPp(k<\mH^{-1},\eta)\simeq \l(\frac{2+\beta}{3+2\beta}\r)^2\mathcal{P}_{\mathcal{R}}(k).
\end{align}
It is clear that the spectrum of $\Phi$ becomes constant at the super-horizon scale and can be directly identified with the inflationary power spectrum up to an equation state-dependent factor. This is well known to be directly imprinted in the CMB temperature anisotropies and gives rise to robust constraints, particularly in the early stage of the inflation dynamics.  
On the contrary, the sub-horizon modes, when entering the horizon during the phase under consideration, can undergo non-trivial evolution, and the spectrum changes according to
\begin{align}
   \mPp(k>\mH^{-1},\eta)\simeq \frac{2^{2(\beta+2)\Gamma^2(\beta+5/2)}}{2\pi}(k\eta)^{-2(\beta+2)}\mathcal{P}_{\mathcal{R}}(k).
\end{align}
The value of $\beta(w)$ assumes the value in between $-1/2\leq \beta\leq 1$. Hence, the comoving power spectrum always decays after the modes' reentry during the post-inflationary epoch. Further, an important point to realize is that CMB temperature anisotropies do not provide any constraint on the spectrum of those sub-horizon modes, and therefore, the later stage of inflationary dynamics remains poorly constrained.  
This particular fact leads to a possibility of modifying the later stage of inflationary dynamics that can modify the scalar power spectrum at small scales with large amplitude \cite{Germani:2017bcs, Martin:2013tda, Leach:2001zf, Kinney:2005vj, Garcia-Bellido:2017mdw, PhysRevD.64.123514, Di:2017ndc, PhysRevD.103.083510, Motohashi:2017kbs, Dimopoulos:2017ged, Fu:2019vqc, Solbi:2021wbo,Figueroa:2021zah}. 
Such enhancement in the curvature spectrum causes large density perturbations and produces primordial black holes (PBH), which is widely discussed as a potential dark matter candidate \cite{PhysRevD.64.123514, Di:2017ndc, PhysRevD.103.083510}. We will later see that those PBH-like power spectra indeed help increase the magnetic field strength on large scales.

In this section, we, therefore, explore those different types of inflationary spectra and their role in modifying the strength of large-scale magnetic fields. We further study the effect of the post-inflationary era, such as reheating with the generic equation of state $\wre$.
Inflation is governed by the slow-rolling inflaton field with a nearly constant potential. 
In such an inflaton background, the curvature power spectrum assumes a power law form. For long-wavelength modes, CMB fixes two parameters of the power law spectrum. However, as we pointed out earlier, a high-frequency curvature power spectrum is poorly constrained, and hence, we parameterized the full inflationary spectrum in the following way~\cite{Cai:2018dig},     
\begin{align}
    \mathcal{P_R}(k) = A_s \left(\frac{k}{k_*}\right)^{n_s-1} + A_0~\text{Exp}\left[-\frac{(k-\kp)^2}{\delta\times \kp^2}\right], \label{eq:PR_2}
\end{align} 
\begin{figure}[t]
\begin{center}
\includegraphics[scale=0.4]{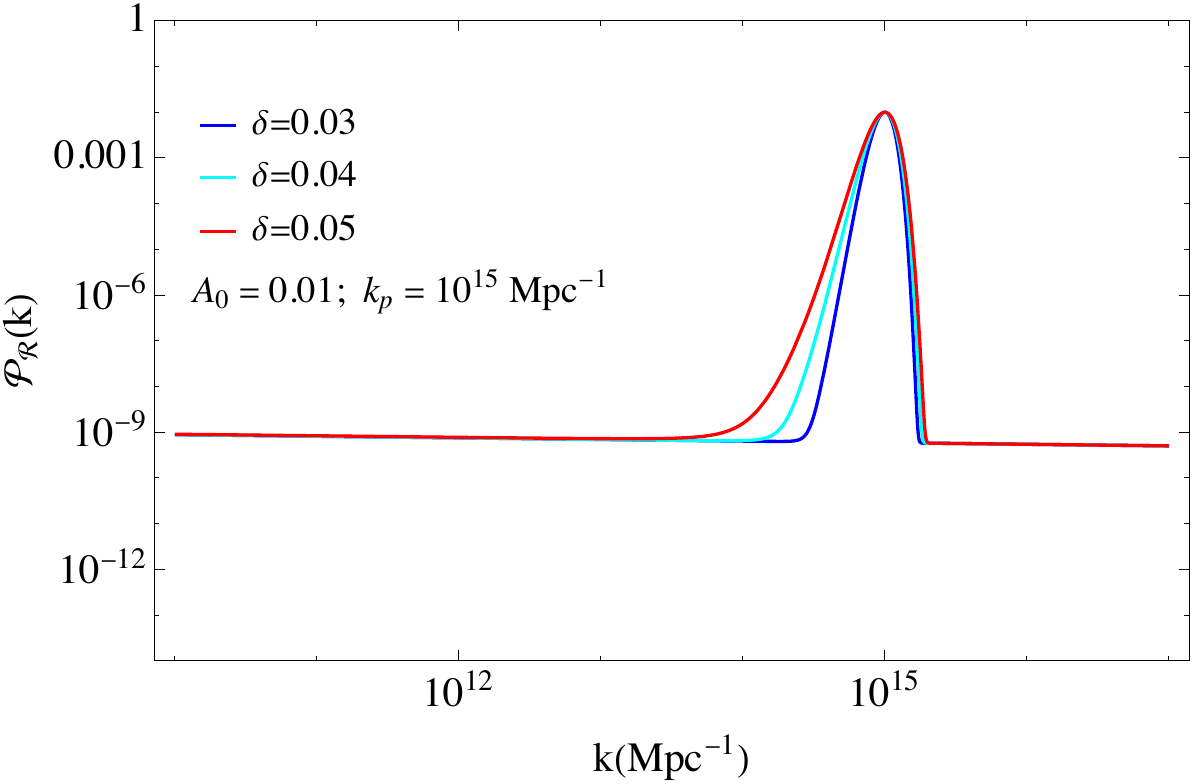}
\includegraphics[scale=0.4]{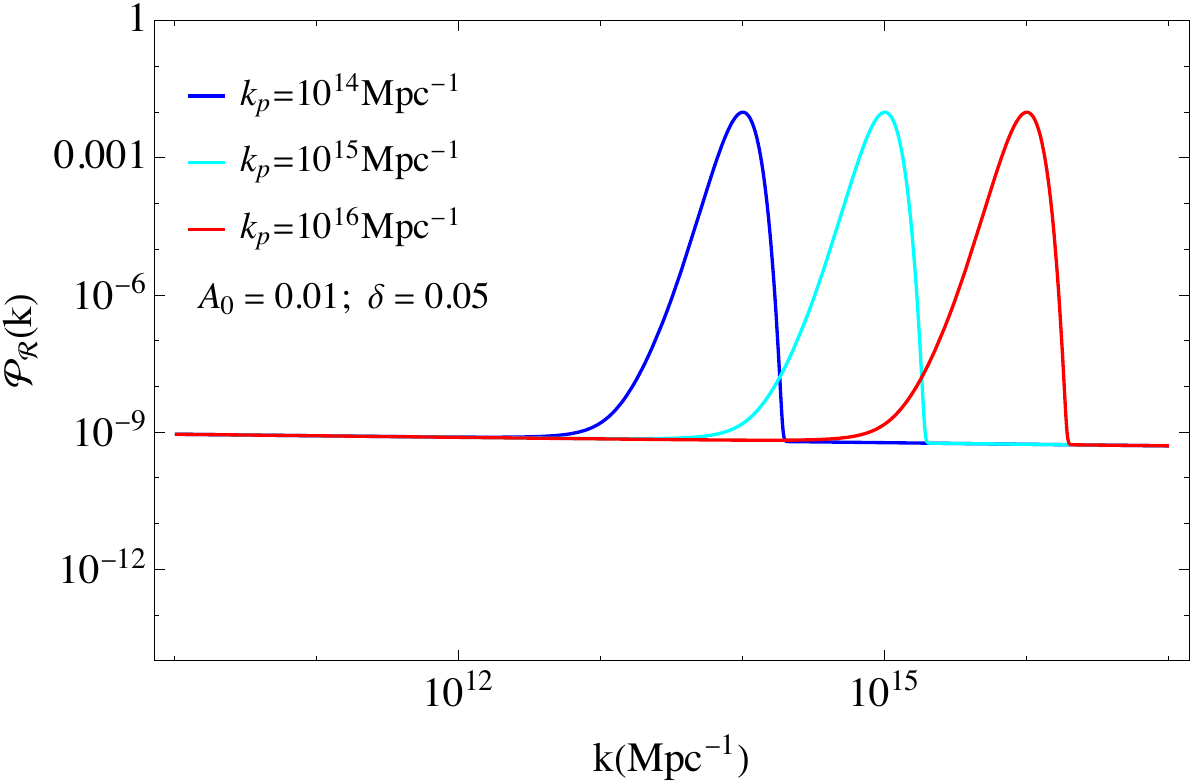}
 \caption{In this figure, we illustrate the behavior of the curvature power spectrum for different scenarios. In the left panel, we show how the spectral shape depends on the parameter $\delta$, represented by three colors. In the right panel, we demonstrate how the spectral peak shifts depending on the peak wave number, denoted as $\kp$, for a fixed value of  $\delta = 0.05$. In both figures, we set the amplitude to $A_0 = 0.01$, the scalar amplitude to  $\mathcal{A}_s \simeq 2.1 \times 10^{-9}$ , and the spectral index to $\ns = 0.965 $.}
    \label{fig_PR}
\end{center}
\end{figure}
 $A_s = 2.1\times 10^{-9}$, and scalar spectral index $n_s=0.9649\pm 0.0042$ at $68\%$ CL
 \cite{Planck:2018vyg}. The second part of the spectrum peaked around $\kp >> k_*$; we parameterize by an exponential function keeping all the essential features of the PBH spectrum studied in the literature \cite{PhysRevD.64.123514, Di:2017ndc, PhysRevD.103.083510, Ragavendra:2020vud, Bhaumik:2020dor, Solbi:2021wbo, Figueroa:2021zah}. This essentially features an exponential amplification of curvature perturbations on small scales. By properly tuning the amplitude parameter $A_0$, one can achieve a magnitude of the power spectrum \(\mathcal{P_R}(k)\) as high as \(\mathcal{O}(10^{-2} - 10^{-1})\). Such a high value is critical for PBH formation in the subsequent evolution. The width of the spectrum is controlled by \(\delta\). To remain consistent with CMB observations, we maintain the following constraint \(\delta^{-1} \leq \ln(A_0 / A_s)\) throughout our analysis.

Utilizing the above curvature spectrum Eq.\ref{eq:PR_2} into the expression of magnetic power spectrum Eq.\ref{magpower}, we find the following approximate analytic expression for the present-day magnetic power spectrum as:  
\begin{align}
    \mathcal{P}_{\rm B}(k, \eta_0)& = \mathcal{P}^{\rm inf}_{\rm B}(k, \eta_e) \l(\frac{\ae}{a_0}\r)^4  \nonumber\\
    &\simeq \frac{8\pi^2\HI^4}{9}  \l(\frac{k}{\ke}\r)^3\l(\frac{\ae}{a_0}\r)^4 \left(\frac{2+\beta}{3+2\beta}\right)^2
    \left[\frac{A_s}{n_s} \l(\frac{\ke}{k_*}\r)^{\ns-1} + \frac{1}{2} \sqrt{\pi\delta} \frac{A_0\kp}{\ke} 
    \left(1 + \text{Erf}\left[\frac{\kp - k}{\kp \sqrt{\delta}}\right]\right)\right] 
    \label{eq:rho_PB_g}
\end{align}  

\begin{figure}[t]
\begin{center}
\includegraphics[scale=0.4]{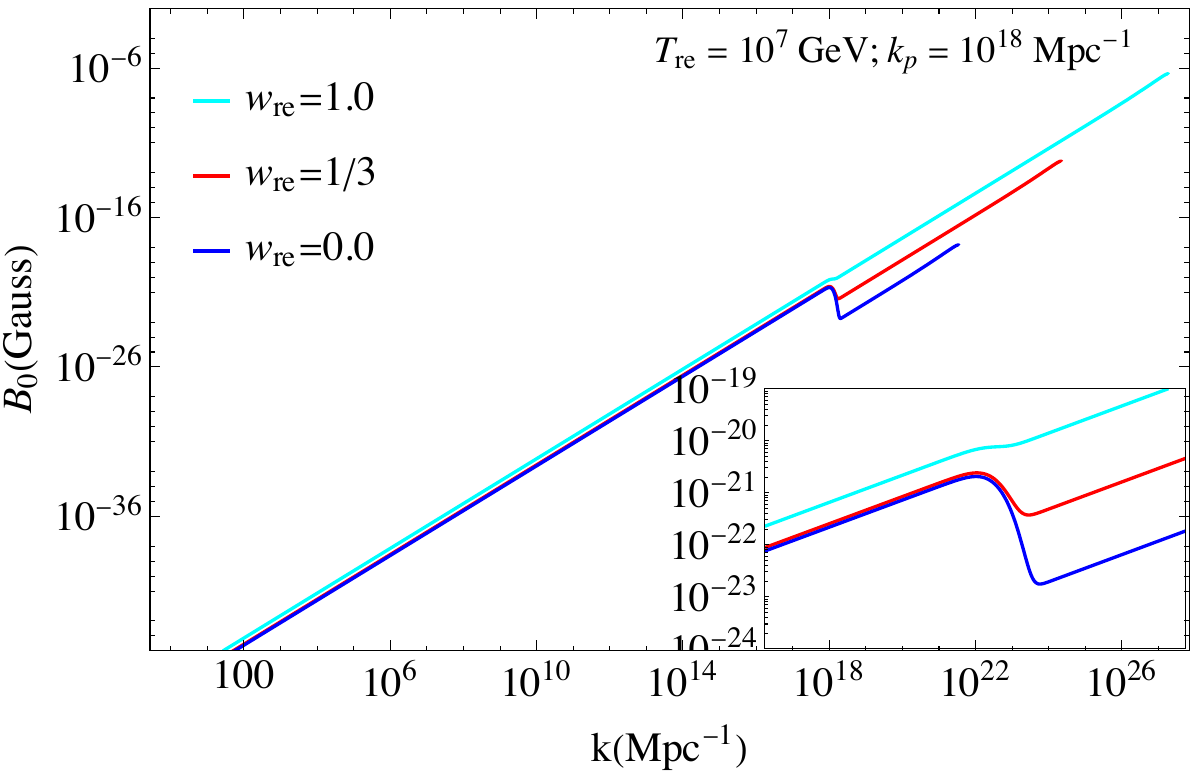}
\includegraphics[scale=0.4]{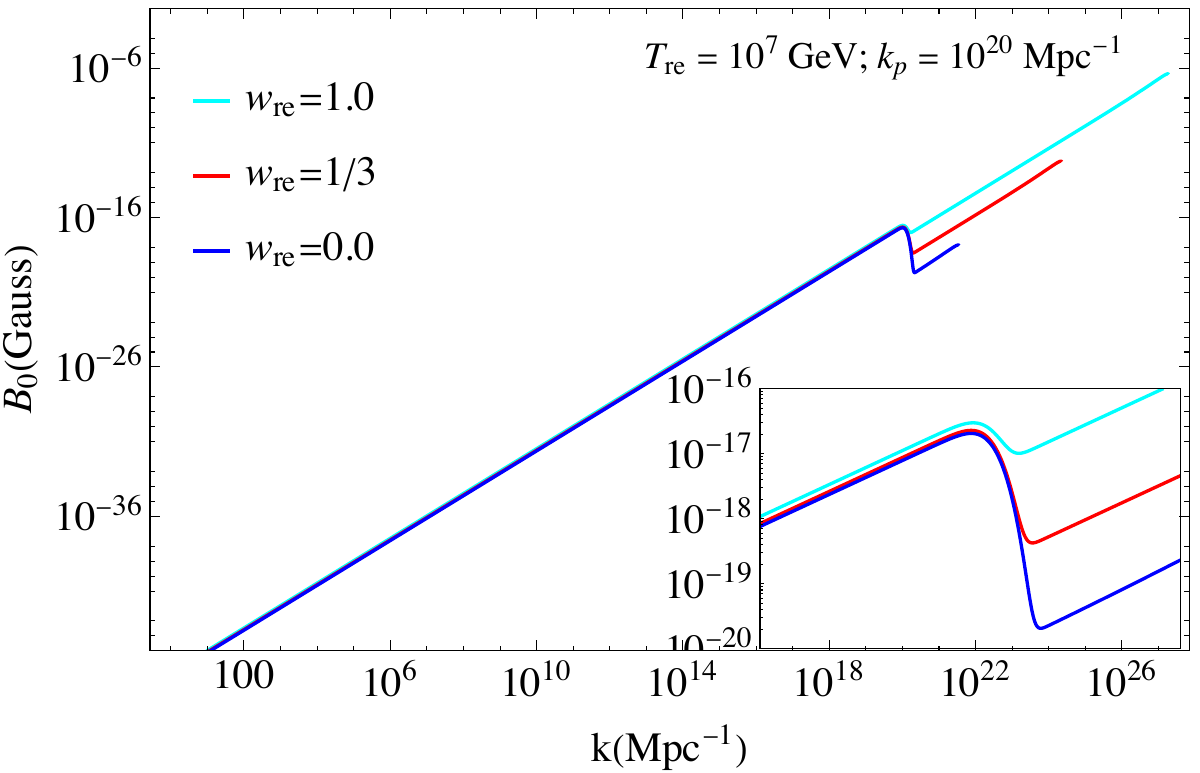}
 \caption{In these figures, we present the present-day magnetic field strength as a function of the comoving wavenumber \( k \) (in \( \Mpc^{-1} \)). We consider a fixed reheating temperature of \( \Tre = 10^7\,\mathrm{GeV} \), with the three different colors corresponding to different values of the equation of state parameter during reheating: \( \wre = 0.0 \), \( \wre = 1/3 \), and \( \wre = 1.0 \). The left panel corresponds to a curvature power spectrum peak at \( \kp = 10^{18}\,\Mpc^{-1} \), while the right panel shows the result for \( \kp = 10^{20}\,\Mpc^{-1} \). In both figures, we set the amplitude of the curvature power spectrum peak to \( A_0 = 0.1 \). For the slow-roll type curvature power spectrum, we use \( A_s \simeq 2.1 \times 10^{-9} \) with a spectral index of \( \ns = 0.965 \).}
    \label{fig_b0_k_w}
\end{center}
\end{figure}


\begin{figure}[t]
\begin{center}
\includegraphics[scale=0.4]{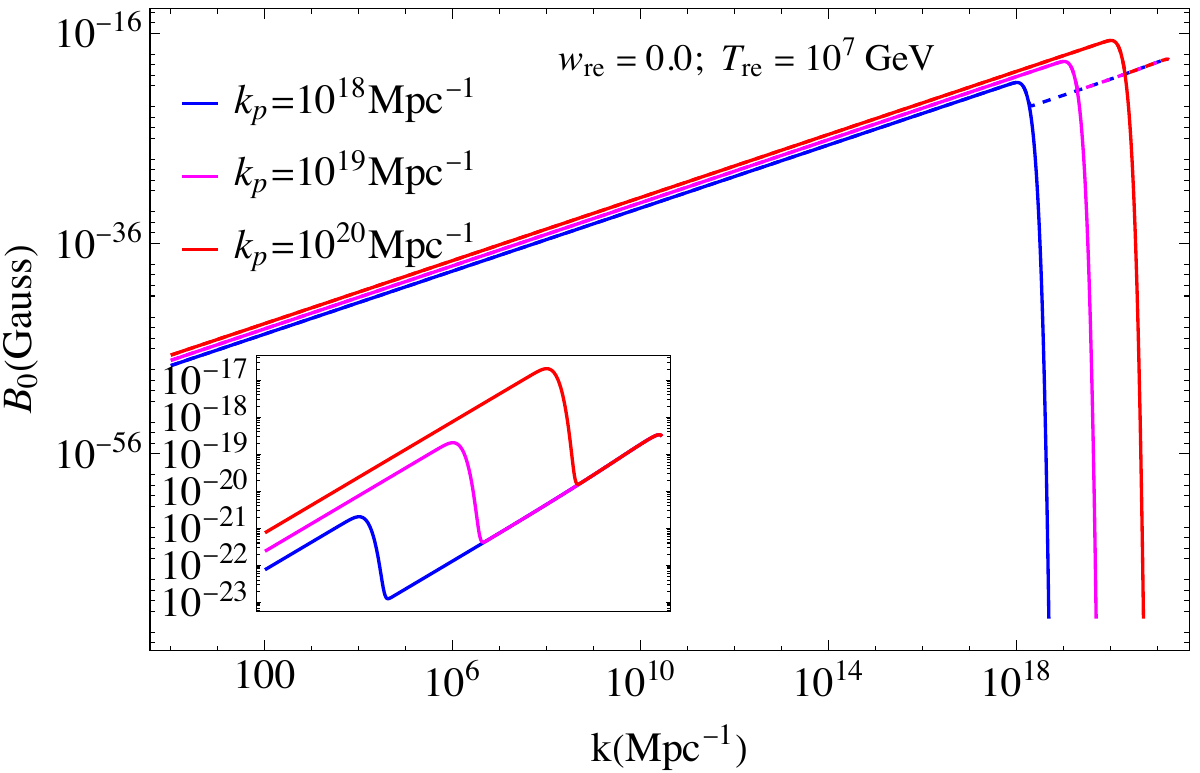}
\includegraphics[scale=0.4]{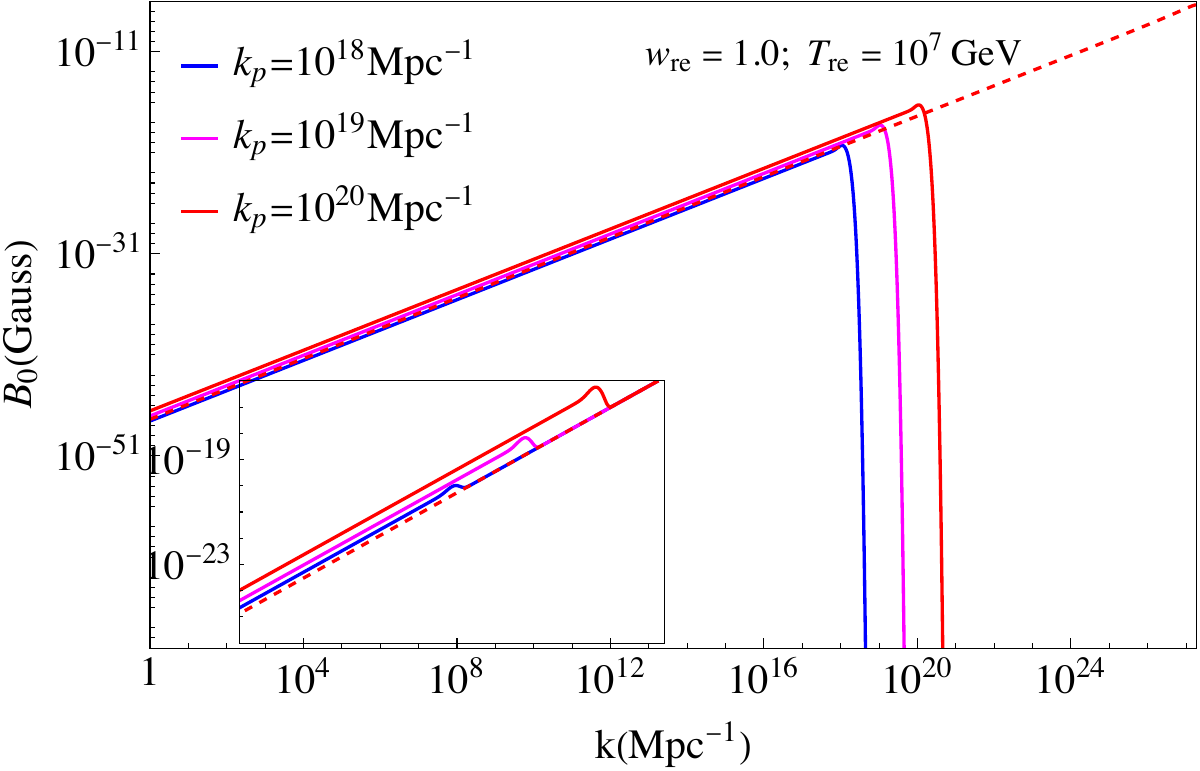}
 \caption{This figures show the present-day magnetic field strength, \( B_0(k) \), as a function of the comoving wavenumber \( k \) (in \( \Mpc^{-1} \)), for two different values of the equation of state parameter during reheating: \( \wre = 0.0 \) (left) and \( \wre = 1.0 \) (right). In both cases, the reheating temperature is fixed at \( \Tre = 10^7\,\Gev \). Each figure includes three curves, with different colors representing three distinct values of the peak scale \( \kp \) of the curvature power spectrum. In both figures, we set the amplitude of the curvature power spectrum peak to \( A_0 = 0.1 \). For the slow-roll type curvature power spectrum, we use \( A_s \simeq 2.1 \times 10^{-9} \) with a spectral index of \( \ns = 0.965 \).}
    \label{fig_b0_kp}
\end{center}
\end{figure}
Here, $(\ae/a_0)$ takes care of the dilution factor of the magnetic energy density due to background expansion from its inflationary counterpart $\mathcal{P}^{\rm inf}_{\rm B}(k, \eta_e)$ defined at the end of inflation.

\begin{figure}[t]
\centering
\includegraphics[scale=0.265]{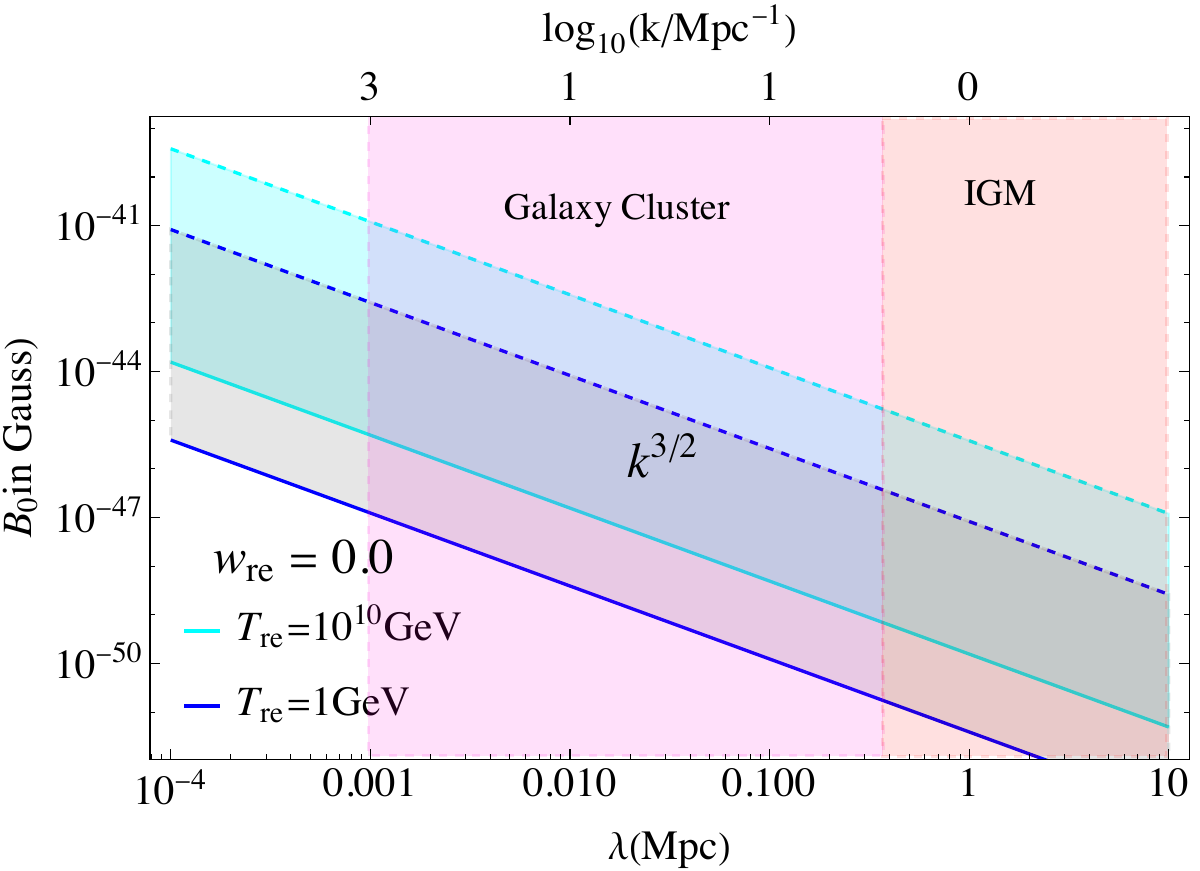}
\includegraphics[scale=0.28]{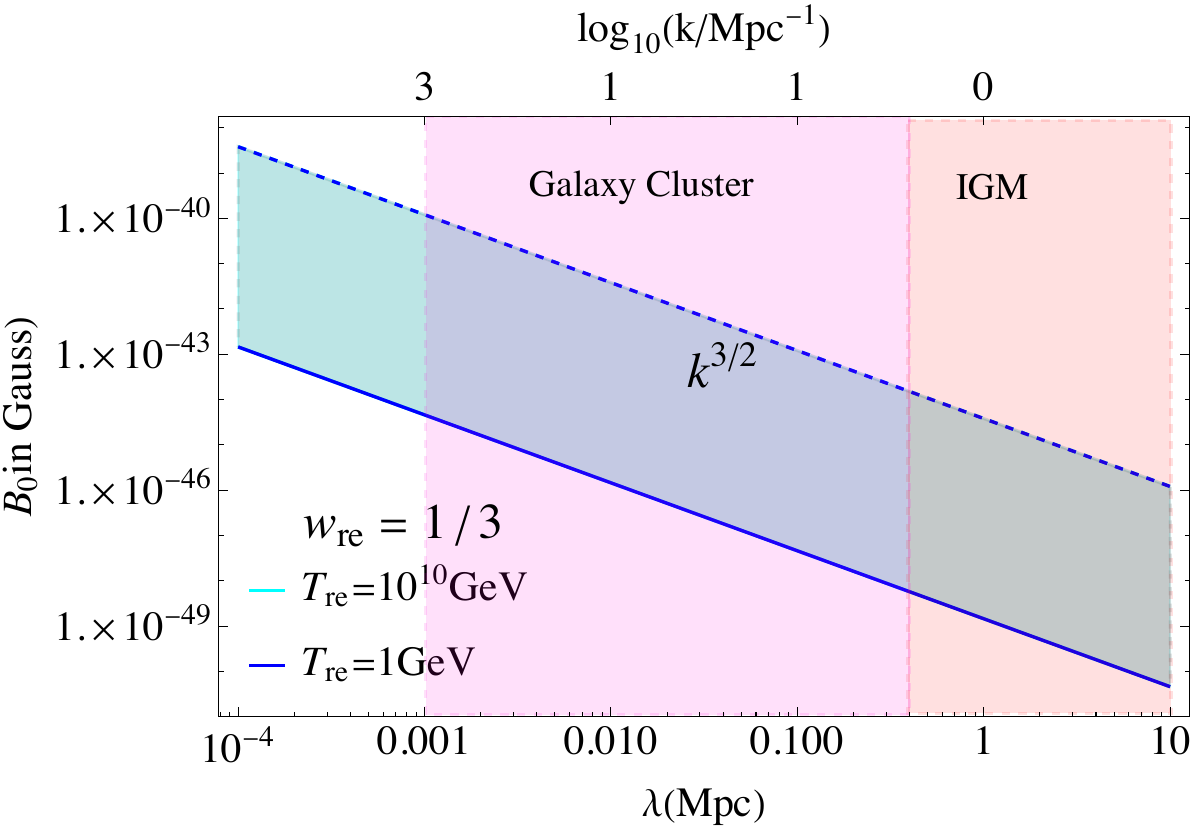}
\includegraphics[scale=0.265]{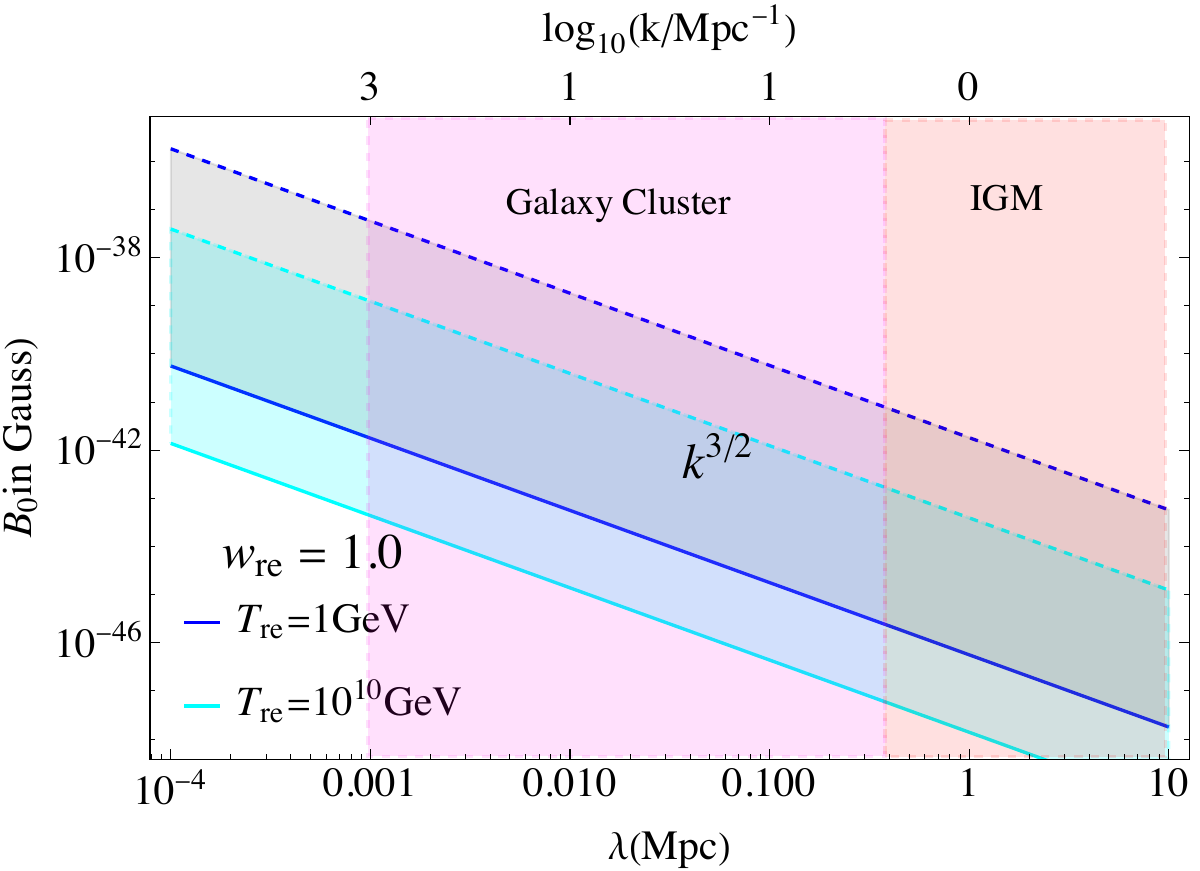}
\caption{The above figures show the present-day magnetic field strength, \( B_0 \) (in Gauss), as a function of the observable wavelength \( \lambda \) (in \( \mathrm{Mpc} \)) for three different values of the reheating equation-of-state parameter: \( \wre = 0.0 \) (left), \( \wre = 1/3 \) (middle), and \( \wre = 1.0 \) (right). The two shaded color bands correspond to two different reheating temperatures: \( \Tre = 1\,\mathrm{GeV} \) (blue) and \( \Tre = 10^{10}\,\mathrm{GeV} \) (cyan). Solid lines represent the magnetic field spectrum generated from a simple slow-roll-type curvature power spectrum, while dashed lines correspond to the PBH-type curvature power spectrum, where the peak is located near the scale \( \ke \), i.e., \( \kp \simeq \ke \). In all plots, we have fixed the maximum amplitude of the curvature power spectrum to be \( A_0 =1.0 \). The vertical shaded regions indicate the characteristic length scales of the intergalactic medium (IGM) and galaxy clusters.
}
\label{fig_b0_lambda}
\end{figure}

Note that $k_* < \kp < k_e$, and the PBH spectrum amplitude $A_0 \gg A_s \sim 10^{-9}$. This immediately suggests that across all scales for $k < \kp$, the magnetic spectrum behaves as \(\mP_{\rm B}(k) \propto A_0 \kp k^3\). The large-scale magnetic field strength is controlled by the amplitude $A_0$ of the small-scale modification of the PBH spectrum and its peak frequency $\kp$. For small wavelength modes, (\(k > \kp\)), as the spectrum amplitude reduces to the primordial one (see Fig.(\ref{fig_PR})), the magnetic spectrum reduces to \(\mP_{\rm B}(k) \propto A_s k^3\) with reduced amplitude. 
In summary, we find the present-day magnetic field strength as a function of the amplitude of the inflationary power spectrum, as,
\begin{equation}\label{Bo}
 B_0 =\sqrt{\mathcal{P}_{\rm B}(k, \eta_0)} \propto
   \begin{cases}
      \sqrt{ A_0 \kp} k^\frac32 \quad &\text{for} \quad k < \kp\\
        \sqrt{ A_s} k^\frac32 \quad &\text{for} \quad k > \kp 
        \end{cases} .
\end{equation}
Whereas near the peak scale, the magnetic field strength shows a fall in strength due to the Error function, and it exhibits a spectral break. However, the peak in the magnetic spectrum near \(k_p\) occurs only if $ \kp\geq {A_s \kp}/{n_s \sqrt{\pi \delta}}$. 
All the aforementioned features observed from the analytic form of the magnetic spectrum are depicted in Figs.~(\ref{fig_b0_k_w}), (\ref{fig_b0_kp}), and (\ref{fig_b0_lambda}). 

At this stage, we provide some numerical estimates to gain insight into the magnetic field strength defined at the comoving scale of \( 1\,\Mpc \). We consider the case where the curvature power spectrum is peaked at \( k \simeq \ke \). Since the post-inflationary dynamics depend on the parameters of the inflationary model, we express all relevant quantities in terms of two reheating parameters: the equation of state (EoS) \( \wre \) and the reheating temperature \( \Tre \) (see Eqs.~\eqref{eq:kre}\eqref{eq:ke} for the relationship between them).
For a matter-dominated reheating phase with \( \wre = 0 \), the magnetic field strength at the end of inflation turned out to be \( B^{\rm inf} = \sqrt{\mathcal{P}^{\rm inf}_{\rm B}(k = 1 \mbox{Mpc}, \eta_e)} \simeq 2.93 \times 10^{16}\, \text{G} \). In the radiation-dominated case with \( \wre ={1}/{3} \), we find \(B^{\rm inf} \simeq 4.29 \times 10^{10}\, \text{G} \), while for a kination-dominated era with \( \wre= 1 \), the magnetic field strength is reduced to \( B^{\rm inf} \simeq 6.25 \times 10^4\, \text{G} \).

Following their production, the magnetic fields dilute as \( a^{-2} \) due to the expansion of the Universe. Consequently, the present-day magnetic field strength depends on the post-inflationary evolution. Agian at 1 Mpc scale for \( \wre = 0 \), we obtain \( B_0 \simeq 3.76 \times 10^{-47}\, \text{G} \); for a radiation-like reheating phase (\( \wre = {1}/{3} \)), \( B_0 \simeq 3.81 \times 10^{-45}\, \text{G} \); and for a kination-like phase (\( \wre = 1 \)), the present-day strength becomes \( B_0 \simeq 3.94 \times 10^{-43}\, \text{G} \). In all these estimates, we assume that the reheating phase ends when the background temperature reaches \( \Tre \simeq 10^4\, \Gev \).

In Fig.\ref{fig_b0_k_w}, we have numerically plotted the present-day magnetic field strength \(B_0(k)\) as a function of the wave number \(k\) for a Gaussian-type curvature power spectrum, as defined in Eq.~\eqref{eq:PR_2}. We consider three sample equations of state (EoS): \(\wre = 0, 1/3, \text{and } 1\). In both panels, the reheating temperature of the universe is fixed at \(\Tre = 10^7~\text{GeV}\). The left panel corresponds to \(k_p = 10^{18}~\text{Mpc}^{-1}\), while the right panel corresponds to \(k_p = 10^{20}~\text{Mpc}^{-1}\). The strength of the magnetic field recovers our analytical expectation $B_0 \propto k^{3/2}$ with a very mild direct dependence on the reheating equation of state.  

As discussed earlier, the cutoff wavenumber \(\ke \) depends on the reheating dynamics and the specific inflationary model under consideration (see Eq.~\eqref{eq:ke}). In general, a higher equation of state parameter (\( \wre > 1/3 \)) corresponds to a shorter duration of the reheating phase compared to a matter-like phase (\( \wre < 1/3 \)) and due to this the produced magnetic field strength is dependent on the duration of the reheating dynamics.
Since the magnetic field generated in such mechanisms is scale-dependent, and the comoving magnetic field amplitude scales as \( B_0 \propto k^{3/2} \), the total magnetic energy density is sensitive to the details of the post-inflationary evolution. From our analysis, we find that while the magnetic field strength at large scales does exhibit dependence on the reheating dynamics, the effect is not significant enough to account for the present-day observed magnetic field strength at cosmological scales, but the total magnetic energy density is strongly dependent on the reheating dynamics.

Similarly, in Fig.~\ref{fig_b0_kp}, we plot the present-day magnetic field \(B_0(k)\) as a function of the comoving wavenumber \(k\) for two different reheating scenarios. The left panel corresponds to \(\wre = 0\) with \(\Tre = 10^7~\text{GeV}\), while the right panel corresponds to \(\wre = 1\) with \(\Tre = 10^7~\Gev\). Three different colors indicate different values of the peak wavenumber: \(k_p = 10^{18}, 10^{19}, \text{and } 10^{20}~\Mpc^{-1}\). 
We observe that the magnetic field strength for scales \(k < k_p\) indeed increases with the peak wave number \(k_p\), following the relation \(B_0(k < k_p) \propto k_p^{1/2}\), as we derived from Eq.~\eqref{Bo}. The dashed lines represent the contribution to the magnetic field from the first term in the curvature power spectrum in Eq.~\eqref{eq:PR_2}, which enables magnetic field generation even at scales \(k > k_p\).

The PBH-induced curvature power spectrum leads to a significant enhancement in the magnetic power spectrum compared to the conventional inflationary scalar power spectrum. Notably, this enhancement extends to all super-horizon modes smaller than the mode $k_p$. From Eq.~\eqref{Bo}, the total enhancement factor from the PBH-type curvature spectrum can be expressed as ${(A_0 k_p n_s)}/{(\ke A_s)}$,
where \(A_0\) is the peak amplitude of the curvature perturbations, \(A_s\) is the scalar amplitude, and \(n_s\) is the spectral index constrained by CMB observations. The maximum enhancement occurs when the peak frequency is aligned with \(\ke\) (i.e., \(k_p \simeq \ke\)), giving an overall enhancement factor of ${(A_0 n_s)}/{A_s}$.
For example, if \(A_0 = 1\) (the maximum allowed curvature amplitude), \(n_s = 0.965\) (the central value from CMB constraints), and \(A_s \simeq 2.2 \times 10^{-9}\), the enhancement factor is approximately \(4.4 \times 10^8\). For purely inflationary spectrum, the magnetic field strength at 1 Mpc scale can be calculated as $\sim 10^{-45}$ Gauss. However, PBH spectrum can enhance that value to a maximum $\sim 10^{-41}$ Gauss. This is indeed a significant enhancement, but much below the current observational bound $\sim 10^{-18}$ Gauss.

\begin{figure}[t]
\centering
\includegraphics[scale=0.5]{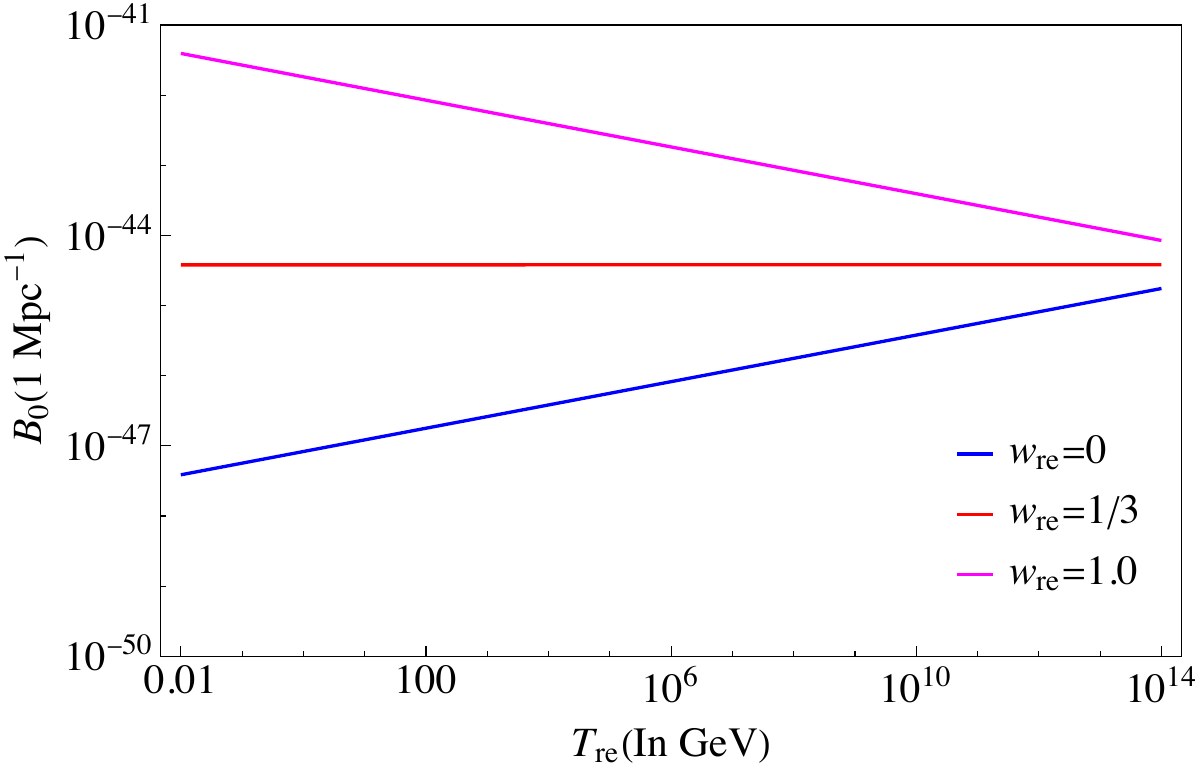}
\caption{In the above figure, we have shown how the present-day magnetic field depends on the reheating temperature (effectively determined by the duration of the reheating periods) for three different values of effective equation of state $\wre=0,\,1/3,~\&\,1.0$ denoted by three different colored lines. Here we consider the peak of the curvature perturbation situated near the $\ke$ to get the maximum enhancement. In all cases, we consider the amplitude of the peak curvature power spectrum to be \( A_0 \simeq 1.0 \) in order to investigate the maximum possible enhancement of the magnetic field due to the curvature power spectrum.}
\label{fig:b0_vs_tre}
\end{figure}

In Fig.~\ref{fig_b0_lambda}, we present the present-day magnetic field strength as a function of wavelength \( \lambda \), expressed in Mpc units. This plot focuses on large-scale magnetic fields, covering a range from \( 10~\Mpc \) to \( 0.1~\mathrm{Kpc} \). We consider three different values of the equation-of-state parameter during reheating: \( \wre = 0 \) (left), \(\wre = 1/3 \) (middle), and \( \wre = 1 \) (right). For each case, we show results for two different reheating temperatures: \( \Tre = 10^{10}\,\Gev \) (cyan) and \( \Tre = 1\,\Gev \) (blue). 
The colored bands represent the range between the maximum and minimum magnetic field strengths generated due to the curvature power spectrum. Solid lines correspond to magnetic fields sourced by the standard slow-roll-type curvature power spectrum, while dashed lines correspond to those sourced by an enhanced curvature power spectrum. To maximize the enhancement, we choose the peak of the curvature spectrum to lie near \( k_p \simeq \ke \), as the amplified magnetic field scales as \( B_0 \propto \ke^{1/2} \).
The shaded vertical regions in the plot indicate different physical length scales. We observe that for higher reheating temperatures, the large-scale magnetic fields sourced by the curvature power spectrum become significantly stronger. In particular, the field strength can be enhanced by up to four orders of magnitude for larger values of \( \wre\).
Notably, in the radiation-like reheating scenario (\( \wre = 1/3 \)), the post-inflationary background evolves like a radiation fluid, making the total duration of the post-inflationary era largely independent of the reheating temperature. In contrast, for matter-like or kination-like reheating scenarios, the duration of the reheating era depends strongly on \( \Tre \), as the background energy density dilutes more slowly (matter-like) or more rapidly (kination-like) compared to the magnetic field energy density. This relative dilution plays a crucial role in determining the final magnetic field strength observed today.
We find that for both \( w_{\mathrm{re}} = 0.0 \) and \( w_{\mathrm{re}} = 1.0 \), the present-day magnetic field strength depends on the duration of the reheating era, whereas for \( w_{\mathrm{re}} = 1/3 \), it does not exhibit such dependence. 
Further, we observe a contrasting behavior in the dependence of magnetic field strength on the reheating temperature. For \( w_{\mathrm{re}} < 1/3 \), the magnetic field strength today increases with increasing reheating temperature (see the left panel of Fig.~\ref{fig_b0_lambda}), while for \( w_{\mathrm{re}} > 1/3 \), the trend reverses: the magnetic field strength decreases with increasing reheating temperature (see the right panel of Fig.~\ref{fig_b0_lambda}).

This opposite behavior can be understood as follows:
\begin{equation}
\mathcal{P}_{\rm B}(k,\eta_0)\propto \frac {a^4_e} {a_0^4} = \frac {a^4_e}{\are^4} \frac{\are^4} {a^4_0} 
\sim e^{-4 N_{re}} \frac {T_0^4}{T_{re}^4} 
= \left(\frac{\pi^2 \gre\Tre^4}{90\HI^2\Mp^2}\right)^{\frac{4}{3(1+\wre)}} \frac {T_0^4}{T_{re}^4} \sim \HI^{-\frac{8}{3(1+\wre)}} \Tre^{\frac{4-12\wre}{3(1+\wre)}}. \label{eq:pb_tre}
\end{equation}
Therefore, for the radiation-like reheating phase, the present-day magnetic field turns out to be independent of reheating temperature. Further for $\wre = 0$, the present day magnetic field $B_0 = \sqrt{\mathcal{P}_{\rm B}(k)}  \propto \Tre^{2/3}$, and for $\wre = 1$, $B_0 \propto \Tre^{-4/3}$. We recover all those behaviors numerically as depicted in Figs.(\ref{fig_b0_lambda})(\ref{fig:b0_vs_tre}).In Fig.~\ref{fig_b0_lambda}, we observe that there are no visible spectral breaks, unlike in Fig.~\ref{fig_b0_k_w} or Fig.~\ref{fig_b0_kp}. This is primarily due to the range of scales considered in this plot. Our focus here is on large-scale magnetic fields and understanding how their present-day strength depends on reheating dynamics. To maximize the enhancement, we assume that the peak of the curvature power spectrum lies at smaller scales, close to \( \ke \). Since the spectral break in the magnetic field spectrum typically appears near the peak of the curvature power spectrum, and this peak lies outside the range of scales shown in Fig.~\ref{fig_b0_lambda}, the spectral break is not visible here. However, if we were to plot the magnetic field spectrum at smaller scales, we would observe similar spectral break features as seen in Fig.~\ref{fig_b0_k_w} and Fig.~\ref{fig_b0_kp}.

In Fig.~(\ref{fig:b0_vs_tre}), we plot the present-day magnetic field strength defined at the comoving scale of \( 1\,\Mpc \) for three different reheating equations of state, \( \wre = 0.0,\, 1/3,\, \text{and } 1.0 \), represented by three distinct colors. As observed, in the case of matter-like reheating (\( \wre = 0 \)), decreasing the reheating temperature \( \Tre \) leads to a corresponding decrease in the present-day magnetic field strength.
 This behavior can be attributed to the fact that, for a matter-like expansion, achieving the same reheating temperature requires a longer duration of reheating. A lower reheating temperature implies a prolonged reheating phase, leading to greater dilution of the magnetic field.

Similarly, for kination-like reheating ( $\wre = 1$ ), where the background dilutes more rapidly, the duration of reheating required to reach a given $\Tre$ is shorter. Consequently, for a fixed reheating temperature, changing the equation of state results in significant variations in the present-day magnetic field strength, as illustrated in Fig.~(\ref{fig:b0_vs_tre}). The explicit dependence of the magnetic field strength on the reheating temperature is given in Eq.\eqref{eq:pb_tre}.

\paragraph{\underline{Magnetic field dominated universe: Bound on $\Tre$:}}  
 So far, we have discussed the generation of a magnetic field from the curvature perturbations. While the magnetic field produced on large scales is generally modest in strength compared to that predicted by alternative mechanisms, the associated total energy density can still be substantial.

The magnetic energy density at conformal time \(\eta\) is given by
\begin{align}
    \rhob(\eta)=\int_{\kpv}^{\ke}\frac{\d k}{k}\,\rhob(k,\eta) = \frac{8\pi^2 \HI^4}{9} \left( \frac{\ae}{a} \right)^4 \left( \frac{2 + \beta}{3 + 2\beta} \right)^2 \left[ \frac{A_s}{n_s} \left( \frac{\ke}{k_*} \right)^{n_s - 1} + \frac{1}{2} \sqrt{\pi \delta} \frac{A_0 k_p}{\ke} \right], \label{eq:rho_b_re}
\end{align}
where \(\HI\) is the Hubble parameter during inflation in conformal time, and \(\ae\) is the scale factor at the end of inflation. In this expression, we have taken the ultraviolet (UV) cutoff for the magnetic field spectrum to be \(\ke\), which represents the highest comoving wavenumber that can be excited by inflationary fluctuations.

During inflation, all perturbation modes originate in the Bunch-Davies vacuum while deeply inside the horizon (i.e., \( k \gg aH \)). As the Universe expands and these modes cross the horizon (i.e., when \( k = aH \)), they undergo a quantum-to-classical transition and their amplitudes become effectively frozen under adiabatic conditions. Since both metric and curvature perturbations arise from inflaton field fluctuations, the largest wavenumber that can be influenced corresponds to the maximum scale \( \ke \), which denotes the highest mode to exit the horizon before the end of inflation.

\color{black}
After their production, magnetic fields evolve adiabatically, diluting as \(a^{-4}\) due to cosmic expansion. In contrast, during reheating, the energy density of the inflaton field—which dominates the background until reheating ends—scales as \(a^{-3(1 +\wre)}\). For inflationary models with a stiff equation of state (\(\wre > 1/3\)) during reheating, the background energy density dilutes faster than the magnetic field energy density. This leads to an effective enhancement of the magnetic field's fractional energy density, scaling as \(a^{3\wre - 1}\).

We define the fractional magnetic energy density at the end of reheating as \(\delta_{\rm B} = \rhob(\eta = \ere) / \rho(\eta = \ere)\), where \(\rho(\eta)\) is the total background energy density. Using Eq.~\eqref{eq:rho_b_re} and the expression for the reheating energy density, we obtain
\begin{align}
    \delta_B(\eta = \eta_{\rm re}) \simeq \frac{8\pi^2}{27} \left( \frac{\HI}{\Mp} \right)^2 \left( \frac{2 + \beta}{3 + 2\beta} \right)^2 \left[ \frac{A_s}{n_s} \left( \frac{\ke}{k_*} \right)^{n_s - 1} + \frac{1}{2} \sqrt{\pi \delta} \frac{A_0 k_p}{\ke} \right] \left( \frac{T_{\rm max}}{\Tre} \right)^{\frac{4(3\wre - 1)}{3(1 + \wre)}}.\label{eq:delta_b}
\end{align}
Here, \(T_{\rm max}\) denotes the maximum temperature of the universe, which depends on the inflationary energy scale. For instance, assuming $\HI\simeq 10^{-5}\Mp$, we obtain $T_{\rm max}\simeq 10^{15}\,\Gev$.

As evident from the expression above, for reheating scenarios with $\wre>1/3$, the fractional energy density increases as the reheating temperature $\Tre$ decreases. Conversely, for $\wre<1/3$, the fractional energy density decreases with decreasing \(T_{\rm re}\). For instance, in the case of kination ($\wre=1$), the scaling becomes $\delta_{\rm B}(T >\Tre) \propto T^{-4/3}$.

\begin{figure}[t]
\centering
\includegraphics[scale=0.5]{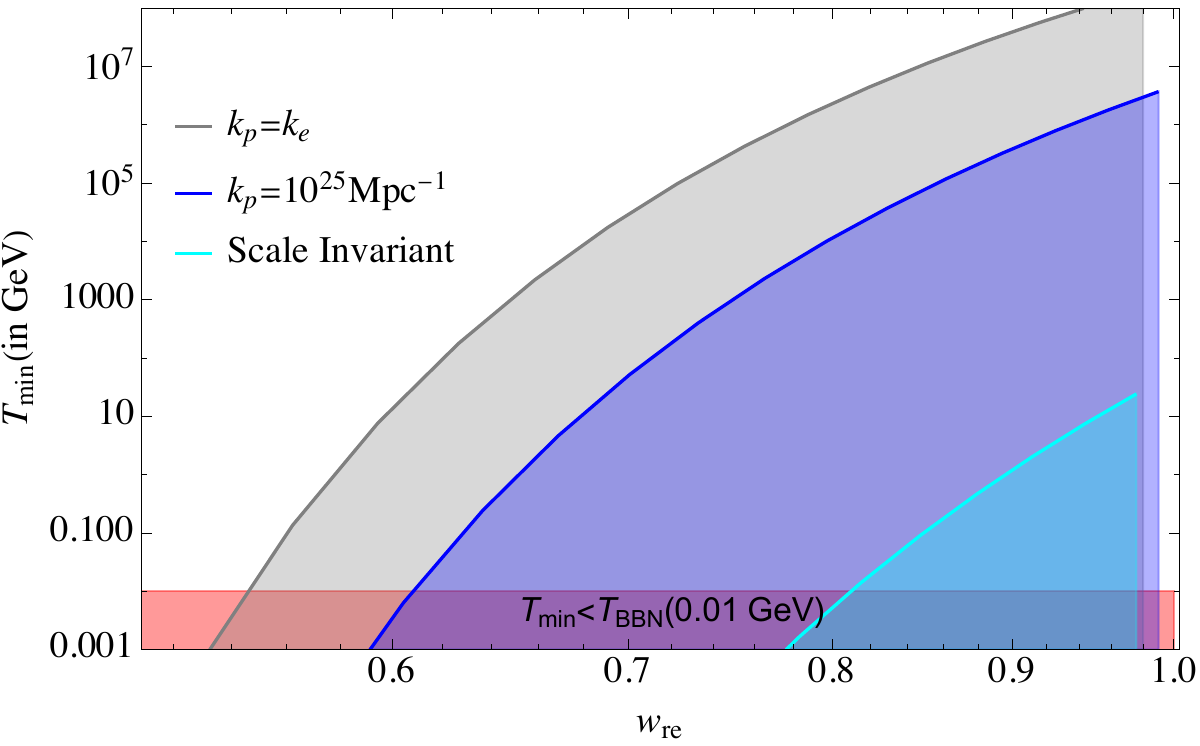}
\caption{In the figure above, we plot the minimum background temperature, \( T_{\rm min} \) (in GeV), as a function of the equation of state parameter during reheating, \( \wre \). The quantity \( T_{\rm min} \) denotes the temperature below which the magnetic energy density exceeds the inflaton energy density, corresponding to the shaded region in the plot. We consider two scenarios for the curvature power spectrum, characterized by different peak scales: the gray curve corresponds to \( k_p = \ke \). Note that $\ke$ is the mode which exits the horizon at the end of inflation. The blue curve represents \( k_p = 10^{25}\, \mathrm{Mpc}^{-1} \). Here the cyan line illustrates the case of a slow-roll type curvature power spectrum. The red-shaded region marks the lower bound on the reheating temperature, given by \( \Tre \simeq T_{\rm BBN} \simeq 0.01\, \mathrm{GeV} \), to ensure consistency with the Big Bang Nucleosynthesis.
}
\label{fig_Tmax}
\end{figure}

To avoid a universe dominated by magnetic fields, the reheating temperature must be bounded from below by the condition \( \delta_B = 1 \). In Fig.~(\ref{fig_Tmax}), we present the minimum allowed reheating temperature as a function of the equation of state parameter \( \wre \), for two different values of the peak scale of the curvature power spectrum. The gray curve corresponds to \( k_p \simeq k_e \), while the blue curve represents \( k_p \simeq 10^{25}\,\mathrm{Mpc}^{-1} \). For comparison, the cyan curve shows the result obtained using a simple slow-roll type curvature power spectrum as defined in Eq.~\eqref{eq:PR_2}. 

We find that for \( \wre = 1.0 \), in order to avoid a magnetic field–dominated universe before the completion of reheating, the minimum reheating temperature must satisfy \( \Tre \gtrsim T_{\rm min} \simeq 4.3 \times 10^8\,\mathrm{GeV} \). In all cases, we consider the amplitude of the peak of the curvature power spectrum to be $A_0\sim0.1$.

Although we have explored the possibility of post-inflationary enhancement of gauge fields, such amplification is constrained by the conductivity of the early Universe. On small scales, it is indeed possible to enhance the curvature power spectrum in scenarios with low conductivity. However, in the presence of a highly conducting plasma, the post-inflationary production of gauge fields sourced by curvature perturbations is significantly suppressed. Even with a primordial black hole (PBH)-type curvature power spectrum, the magnetic field can at most be amplified by a factor of \(\sim 10^5\), which remains far below current observational limits.

Given that our primary objective is to account for the large-scale magnetic fields with coherence lengths larger than \(1~\mathrm{Mpc}\), this post-inflationary mechanism alone is insufficient. Additional dynamics are required to amplify magnetic fields on these large scales. In the following, we discuss how ultra-light axion-like particles (ALPs) can lead to further enhancement of the gauge field, particularly during the Dark Ages. During this epoch, ALPs re-enter the horizon and begin to oscillate, sourcing gauge fields efficiently through their coupling to photons.

It is important to note that post-inflationary amplification due to curvature perturbation is not strictly necessary for the secondary stage of gauge-field production. The same present-day magnetic field spectrum can be achieved even in the case of a simple slow-roll curvature power spectrum, provided that the axion-photon coupling constant is sufficiently large.
\color{black}

\section{Second stage: {\rm Resonant amplification from oscillating ALPs after photon decoupling}}\label{sec_III}
In our previous section, we observed that inhomogeneous metric perturbation during inflation generates a large-scale magnetic field. However, as it is a second-order effect in perturbation, the strength turned out to be insufficient for the current observed bound. In this section, we therefore explore the second phase of gauge field production via ultralight Axion or Axion-like particles (ALPs). Axion is a pseudo-Nambu-Goldstone boson, which can naturally be very light due to the underlying shift symmetry of the system. It couples with the gauge field via the parity-violating coupling $\mfa\, F {\bar F}$ term. Let us suppose the axion is initially misaligned from its true vacuum. In that case, it can dynamically evolve into an oscillatory state, and the coupled gauge field can be copiously produced if appropriate resonance conditions are satisfied. In this section, we will utilize this mechanism in the cosmological background and explore the possibility of resonant magnetic field production at a very large scale. 

However, it is important to realize that after the inflation, the universe becomes filled with highly conducting plasma, and hence, any non-vanishing electric field propagating through it quickly decays due to the rapid response of free charges, and the magnetic field freezes into a constant value. The decay time scale is closely tied to the characteristic frequency of the plasma, which determines the interplay between the charge density oscillations and field dynamics. Any electric field mode frequency larger than the plasma frequency expressed as $\omega_p=(4\pi q^2 n_q/m_q )^{1/2}$ quickly decays within the time scale $\tau\propto \omega_p^{-1}$. Here, $n_q$ is the number density of particles of charge $q$, and $m_q$ is its mass. In the post-inflationary universe, the charge density is expected to be very large; therefore, production of the gauge field from any external source, such as an oscillating axion, may always seem to remain suppressed due to high conductivity.

Thanks to the recombination era in the early universe, the phase after the recombination, well known as the Dark Ages of the universe, comes to our rescue. Recombination is the phase during which neutral hydrogen and Helium are formed, and the plasma frequency becomes negligibly small. This significantly reduces the decay timescale so that the gauge field can no longer feel the conductivity. 
It is in this era that we will discuss the production of gauge fields due to axion-photon coupling.
The Lagrangian density of the axion-photon system is given by
\begin{align}
    \mL =- \frac{1}{2}\partial_\mu\mfa \partial^\mu \mfa -V(\mfa)- \frac{1}{4}\Fmunu\Fmunut -\frac{\alpha}{4\fa} \mfa\, \Fmunu \tilde{F}^{\mu\nu}
\end{align}
where $\alpha$ is the dimensionless coupling parameter and $\fa$ is the decay constant of the axion field denoted as `$\mfa$'. Here $V(\mfa)$ is the potential of the axion (or ALPs) fields. The potential for axion-like particles (ALPs) is typically modeled as \begin{align}
    V(\mfa)=(\ma\, \fa)^2\l[ 1-\cos\l( \frac{\mfa}{\fa}\r)\r].
\end{align}
Here, $\ma$ is the mass of the axion field. 
The axion is assumed to be a purely classical homogeneous field that rolls down to the minima of its potential. 
The equation of motion (EoM) for the homogeneous mode of the axion field $\mfa$ and the Fourier modes of the gauge field $A_\lambda$ in the FLRW background are given by\cite{Patel:2019isj, brandenberger2025}
\begin{align}
    \mfa''+2\mH\mfa'+a^2V_{,\mfa}=\frac{a^2\alpha}{\fa}\langle \bf{E\cdot B}\rangle\label{eq_chik}, \\
    A_\lambda''(k,\eta)+\l(k^2-\lambda \frac{\alpha k\mfa'}{\fa}\r)A_\lambda(k,\eta )=0\label{eq_Ak},\end{align}
Here, \( \lambda \) denotes the polarization index of the gauge field. It is important to note that the above two equations are valid only in the absence of photon scattering, and hence they are applicable during the post-recombination epoch up to reionization.
 The term on the right-hand side of the first equation, which describes the backreaction of a gauge field production, can be expressed as \cite{Patel:2019isj}
\begin{align}
    \langle {\bf E\cdot B}\rangle=-\frac{1}{8\pi^2}\int \frac{dk}{k}\l(\frac{k}{a}\r)^4 \sum_{\lambda=\pm}\l[\frac{1}{k}\frac{d}{d\eta}\l(|\sqrt{2k}A_\lambda|^2\r) \r ]
\end{align}
Nevertheless, we will work in the parameter regime where such backreaction is small. A detailed study will be done in our future publication.  
From the homogeneous axion equation, one can quickly realize that when the dynamical Hubble parameter becomes close to the axion mass, the axion oscillates around its minimum, and we define the associated scale to
as $\kosc=\aosc\Hosc$. Since the universe is matter-dominated by this time, we also see the scale factor evolving during this period as  
$a(\eta)=\aosc(\eta/\eosc)^2=\aosc(x/\xosc)^2$. In terms of the dimensionaless time $x=\kosc\eta$, and field variable $\psi=\mfa/\fa$, we rewrite the above equation as ,
\begin{subequations}
    \begin{align}
  \frac{\partial^2 \psi}{\partial x^2}+\frac{4}{x}\frac{\partial\psi}{\partial x}+\frac{m^2_a}{\Hosc^2}\left(\frac{x}{x_{\rm osc}}\right)^4 \sin \psi &=0 ,\\
     \frac{\partial^2 A_{\lambda}}{\partial x^2}+ \l(\frac{k^2}{k_{\rm osc}^2} -\lambda \alpha \frac{k}{k_{\rm osc}}\frac{\partial\psi}{\partial x}\r)A_{\lambda} &=0 .\label{eq:Ak_1}
\end{align}
\end{subequations}
\begin{figure}[t]
\centering
\includegraphics[scale=0.37]{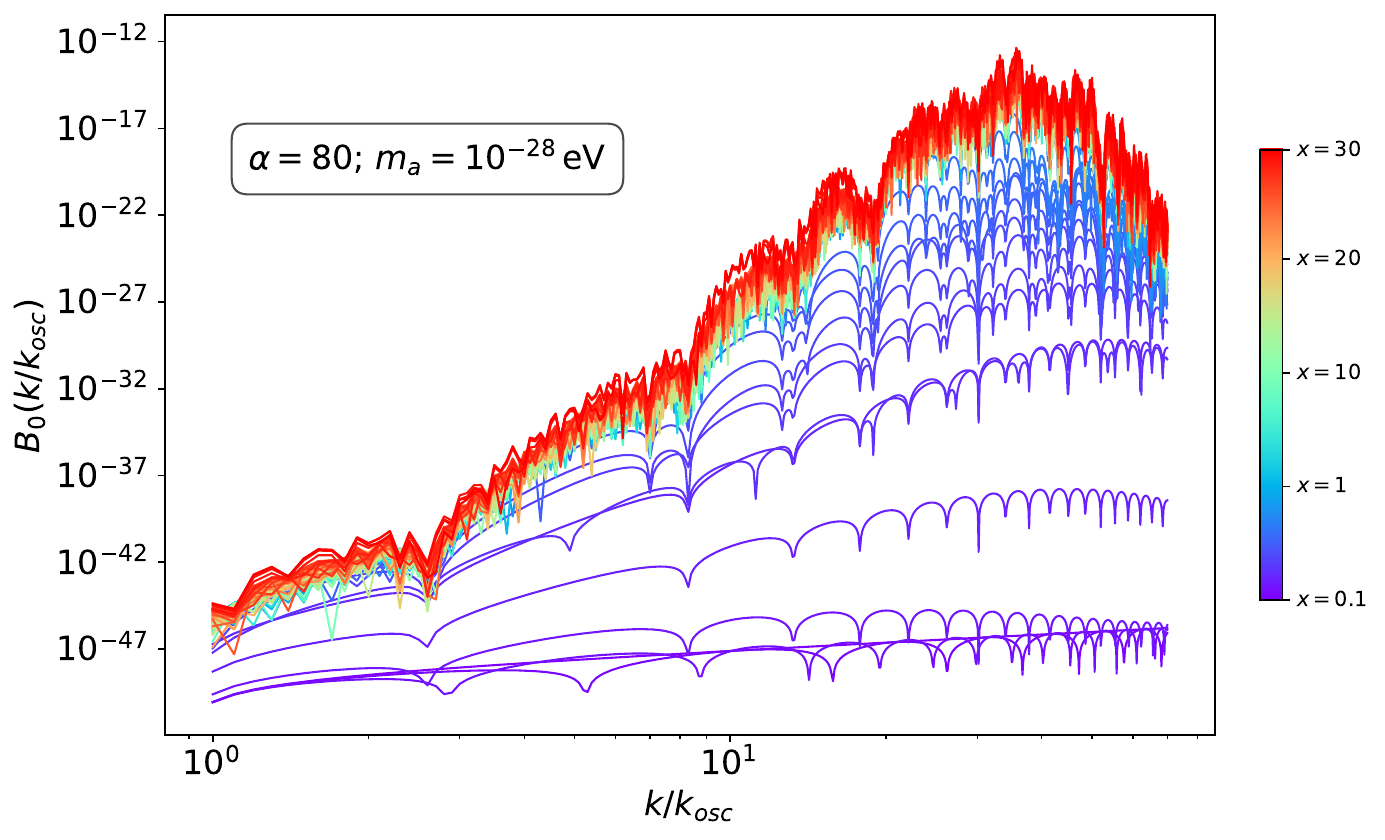}
\includegraphics[scale=0.4]{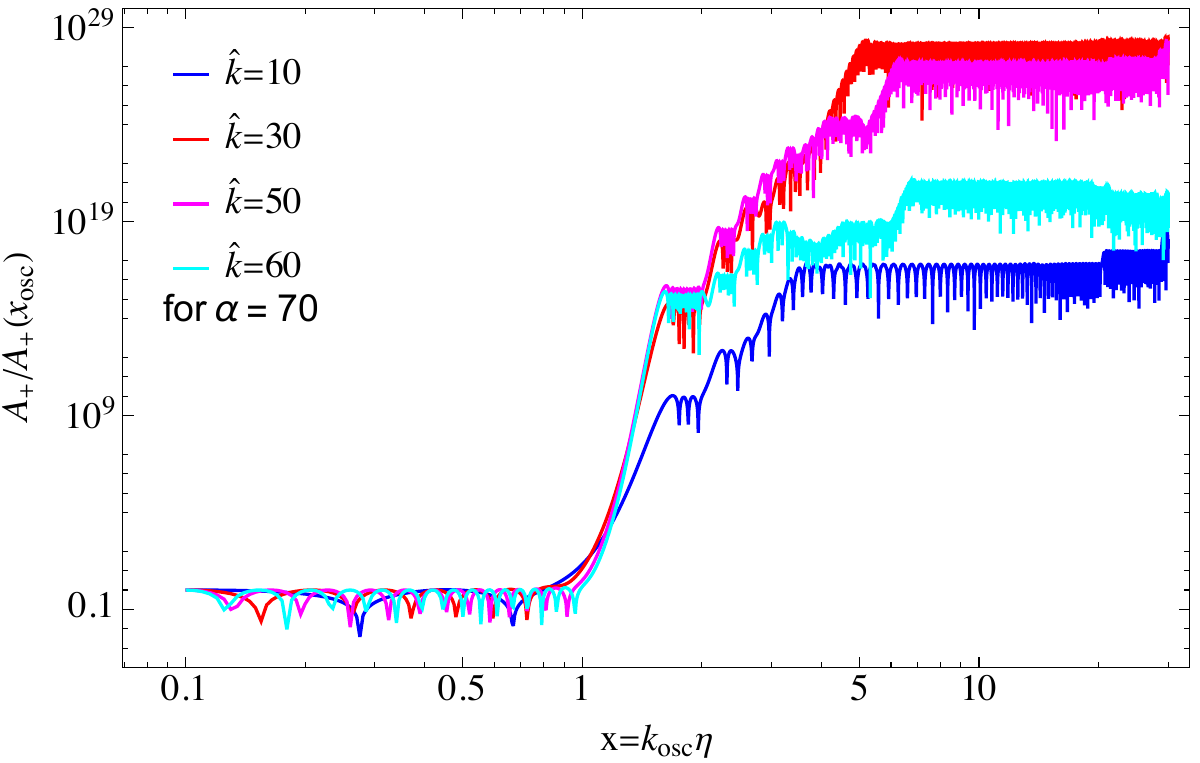}
 \caption{In the left panel, we present the time evolution of the comoving magnetic field as a function of the comoving wavenumber normalized by \(\kosc\), for a fixed value of the coupling \(\alpha = 80\) and axion mass \(\ma = 10^{-28}\,\eV\). The side color bar indicates the time evolution of the comoving magnetic field, starting from \(x = \kosc \eta \simeq 0.1\) and evolving up to \(x \simeq 30\). At the initial time \(x \simeq 0.1\), all relevant wavenumbers lie outside the horizon. The initial condition for the gauge field is set based on the inflationary production discussed earlier, which provides the seed configuration for subsequent evolution. When we set the initial magnetic field strength, then we simply consider a slow-roll type inflationary curvature power spectrum with $A_s\simeq 2.1\times 10^{-9}$ with $\ns\simeq 0.965$.
In the right panel, we show the late-time growth of the gauge field, normalized by its initial value, as a function of the dimensionless time variable \( x = k_{\rm osc} \eta \). This plot corresponds to \( \alpha = 70 \), and we define the dimensionless wavenumber \( \hat{k} = k / \kosc \). }
\label{fig_spc}
\end{figure}

As stated earlier, we assume the axion to be frozen in at $\psi=1.0$ and $\dot{\psi}=0$ until the Hubble parameter becomes close to its mass after the recombination era. Once the condition $m_\chi \simeq 3H$ is met, the axion starts to oscillate in a neutral matter-dominated phase, leading to tachyonic enhancement of the gauge field production. 
We should remember that such production can continue to occur until the universe becomes reionized. 
Utilizing the entropy conservation, the oscillation scale at present can be easily calculated to be, 
\begin{align}
    \frac{\kosc}{a_0}\simeq \l(\frac{43\gs(\Tosc)}{11}\r)^{1/3}\frac{T_0 \ma}{3\Tosc}\label{eq_kosc} \simeq 0.01\,\Mpc^{-1}\,\l(\frac{\ma}{10^{-28}\,\eV}\r) ,
\end{align}
where $\Tosc$ is the temperature of the thermal bath when the axion starts to oscillate, and we can express this as a function of axion mass $m_a$ as
\begin{align}
    \Tosc=\l(\frac{10}{\gs(\Teq)}\frac{\ma^2\Mp^2}{\Teq}\r)^{1/3}\simeq 0.28\,\eV\l(\frac{\ma}{10^{-28}\,\eV}\r)^{1/3}\label{eq_Tosc} 
\end{align}
Here, \( g_*(T_{\mathrm{eq}})\simeq\gs(\Tosc) \simeq 3.36 \) represents the number of relativistic degrees of freedom at the time of matter-radiation equality with the radiation temperature defined as $T_{\rm eq} \sim 0.8$ eV. Note the temperature of the radiation at the time of photon decoupling is defined as \( T_{\rm rec} \simeq 3000 {\rm K} \sim 0.26~\mathrm{eV} \), and at the present time is defined as $T_0 = 2.7$ K = $2.348 \times 10^{-4}$ eV. As pointed out earlier, after this, the universe became dominated by non-relativistic matter, mostly with neutral hydrogen. However, subsequently, cosmic reionization began when the first stars formed, and early galaxies and quasars ionized most of the intergalactic medium. The temperature of the background radiation at the onset of reionization can be estimated to be in the range \( T_{\rm reion} \simeq 0.002 - 0.005~\mathrm{eV} \). And the time between these two distinct phases is known as the Dark Ages. 
As we are particularly interested in the production of gauge fields during the Dark Ages, the axion mass is required to lie within the range $10^{-28} \text{--} 10^{-31}~\eV$. The corresponding background temperatures at which the axion field begins to oscillate are given by $\Tosc(\ma = 10^{-28}\,\eV) \sim 0.28\,\eV$ and $\Tosc(\ma = 10^{-31}\,\eV) \sim 0.003\,\eV$, respectively.
 Utilizing Eqs.~\eqref{eq_kosc} and \eqref{eq_Tosc}, we can compute the corresponding wavenumbers of modes that undergo excitation due to axion oscillations. Our analysis indicates that during the Dark Ages, characteristic length scales in the range \( 1~\mathrm{Mpc} \) to \( 10^3~\mathrm{Mpc} \) can be excited.

\begin{figure}[t]
\centering
\includegraphics[scale=0.4]{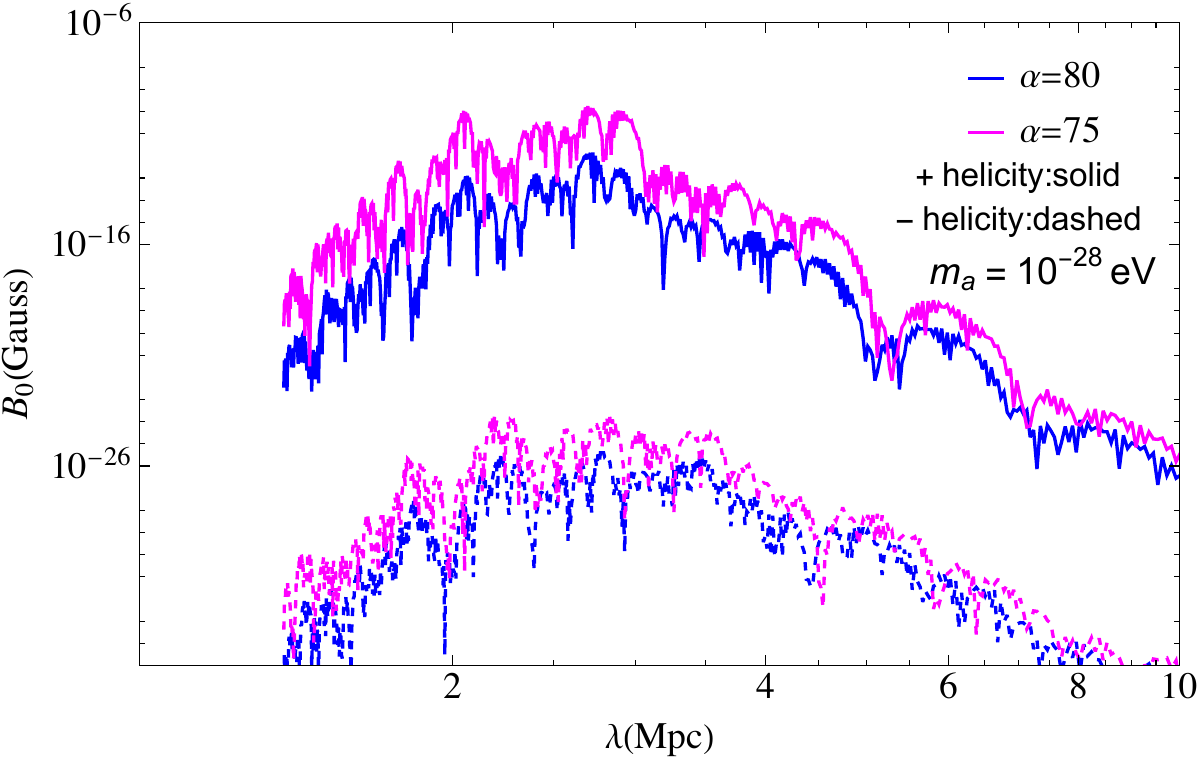}
\includegraphics[scale=0.4]{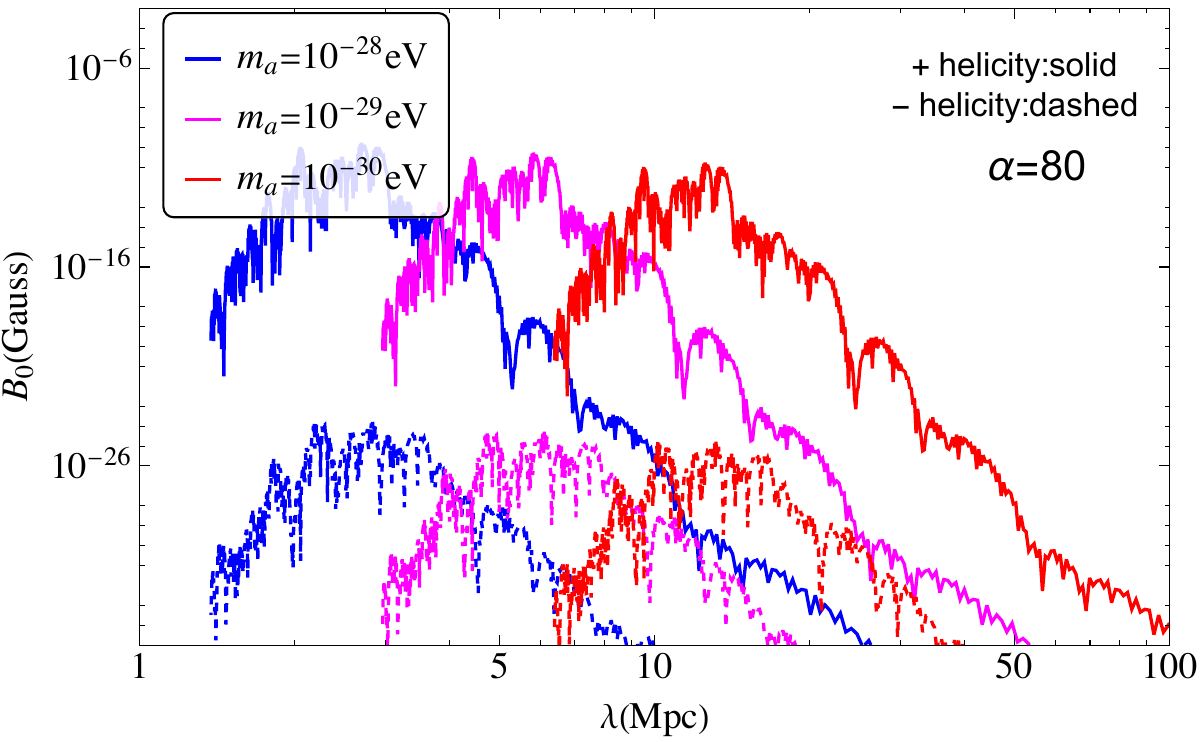}
\caption{Present-day magnetic field strength \( B_0 \) (in Gauss unit) as a function of the comoving length scale \( \lambda \) (in Mpc). \textbf{Left panel:} Variation of \( B_0 \) for two different values of the coupling constant, \( \alpha = 75 \) (blue) and \( \alpha = 80 \) (magenta), with a fixed axion mass \(\ma = 10^{-28}\,\text{eV} \). \textbf{Right panel:} Dependence of \( B_0 \) on the axion mass, shown for \( m_a = 10^{-28} \,\text{eV} \) (blue), \( 10^{-29}\,\text{eV} \) (magenta) and \( 10^{-30}\,\text{eV} \) (red), with a fixed coupling constant \( \alpha = 80 \). In both panels, solid lines represent positive-helicity $(+)$ modes, while dashed lines correspond to negative-helicity $(-)$ modes. For both spectra, we assume that the initial magnetic fields were generated during inflation due to a slow-roll-type curvature power spectrum, characterized by $A_s\simeq 2.1\times 10^{-9}$ and $\ns\simeq0.965$.}
    \label{fig:B0_vs_lambda}
\end{figure}

\paragraph{\underline{Analytical Spectral behaviors of Magnetic fields:}\\}
We can rewrite the EoM of the gauge field Eq.\eqref{eq:Ak_1} in the following fashion
\begin{align}
    \frac{\partial^2A_\lambda}{\partial x^2}+\kappa^2A_\lambda=0, ~~~\mbox{with}~~~\kappa = \sqrt{\l(\frac{k^2}{k_{\rm osc}^2} -\lambda \alpha \frac{k}{k_{\rm osc}}\frac{\partial\psi}{\partial x}\r)}
\end{align}
Each mode evolves with an effective, time-dependent frequency denoted by \(\kappa\). This implies that modes with long wavelengths, those satisfying the condition \(k/k_{\rm osc} < |\lambda \alpha \psi'(x)|\), can experience exponential growth. During a matter-dominated era, the axion field \(\psi(x)\) evolves as $\psi(x) = \psi_0 \,{\sin(x^3)}/{x^3}$, and consequently $\psi'(x) \sim 3 \psi_0 \cos(x^3)/x$.
Substituting this into the instability condition, one finds the mode $k_{\rm max} \sim (3\,\alpha \psi_0 k_{\rm osc})/{x_{\rm osc}}$,
which represents the maximum wavenumber susceptible to the instability induced by axion oscillations. Modes with \(\kosc < k < k_{\rm max}\) are unstable and can grow due to the tachyonic instability, although this growth is limited to a finite range of frequencies determined by the dynamics of the background expansion. Employing the WKB approximation, the gauge field solution in the tachyonic regime, where \(\kappa^2 < 0\), is given by
\begin{align}
    A_\lambda(x)=\frac{C_1}{\sqrt{|\kappa(x)}|} e^{\beta_c(x)}
      ~~~~\mbox{where}~~~~\beta_c(x) 
    = \int_{x_{\rm osc}}^x \sqrt{\l(\frac{k^2}{k_{\rm osc}^2}\r) -\alpha \frac{k}{k_{\rm osc}}\frac{\partial\psi}{\partial x'}}dx'
\end{align}
As $\kappa>0$, the last term of the above equation is oscillatory in nature. Note that axion is frozen until $x_{\rm osc}$, we can safely set the the initial condition at $x=x_{\rm osc}$, $A(x=x_{\rm osc})=A_{\rm osc}$, where $A_{\rm osc}$ is the initial amplitude of the gauge field produced due to the inflationary curvature perturbation discussed in the beginning. Now, utilizing the initial condition, we get the approximate solution of the gauge field as
\begin{align}
    A_\lambda(x)=A_{\rm osc}(k)\l(\frac{|\kappa(x_{\rm osc})|}{|\kappa(x)|}\r)^{1/2}e^{\beta_c(x)}\label{eq:A_aprox}
\end{align}
Here, \( \beta_c \) quantifies the enhancement of the gauge field during axion oscillations. It is important to note that the two polarization modes, \( \lambda = \pm \), experience instability at different times. As a result, the produced electromagnetic field acquires a helical nature.

Now, utilizing the Eq.\eqref{eq:rho_PB_g} and Eq.\eqref{eq:A_aprox}, we can immediately compute the spectral behavior of the present-day magnetic field at large scale as
\begin{equation}
B_0 =\sqrt{ {\cal P}_{B_0}(k)} \propto k^2 A_\lambda(x_0) \propto k^{3/2} e^{ \beta_c(k,\lambda,\alpha)}  
\end{equation}
This is one of the main results of the present paper.
We can also estimate the maximum amplitude of the produced magnetic field by finding the maximum value of $\beta_c$.

Fig.~(\ref{fig_spc}) illustrates the evolution of the generated magnetic field. The left panel shows the time evolution of the comoving magnetic field spectrum as a function of \( k/\kosc \), for a coupling constant \( \alpha = 80 \). In this analysis, the axion mass is fixed at \( m_a = 10^{-28}\,\mathrm{eV} \). The color bar represents the dimensionless time variable \( x \). As time progresses, we observe that the magnetic field spectrum initially grows and eventually saturates, indicating the end of significant amplification.
The right panel shows the time evolution of individual magnetic field modes as a function of the dimensionless time variable \( x = k_{\rm osc}\eta \), for a fixed coupling parameter \( \alpha = 70 \). It indicates the resonant growth of the gauge field in the oscillating axion background. Once the axion oscillation amplitude reduces sufficiently, the growth ceases to exist. 

To study the spectral behavior of the magnetic field, we plot the present-day field strength \( B_0(\lambda) \) as a function of comoving wavelength \( \lambda \) (in Mpc) in Fig.~(\ref{fig:B0_vs_lambda}). The left panel shows results for two values of the coupling parameter: \( \alpha = 80 \) (blue) and \( \alpha = 75 \) (magenta). Solid lines represent the positive helicity modes (\( A_{+} \)), while dashed lines indicate the negative helicity modes (\( A_{-} \)). For this plot, we fix the axion-like particle (ALP) mass at \( m_a = 10^{-28}\,\mathrm{eV} \). As expected, a stronger coupling (\( \alpha \)) leads to greater magnetic field amplification. We also observe a difference between the two helicity modes. This is because each mode experiences instability at different times, leading to different amplification levels.

The right panel of Fig.~(\ref{fig:B0_vs_lambda}) explores how the magnetic field spectrum changes with the ALP mass. Here, we vary the mass over three values: \( m_a = 10^{-28},\,10^{-29},\,\text{and}\,10^{-30}\,\eV \), while keeping the coupling fixed at \( \alpha = 80 \). As before, solid lines correspond to \( A_{+} \) modes, and dashed lines to \( A_{-} \) modes. We find that lighter ALPs shift the peak of the magnetic field spectrum to larger length scales. This shift occurs because lighter ALPs start oscillating later in cosmic history, when longer wavelengths have already entered the horizon. As a result, gauge fields with larger wavelengths are excited more efficiently..

\begin{figure}[t]
\centering
\includegraphics[scale=0.5]{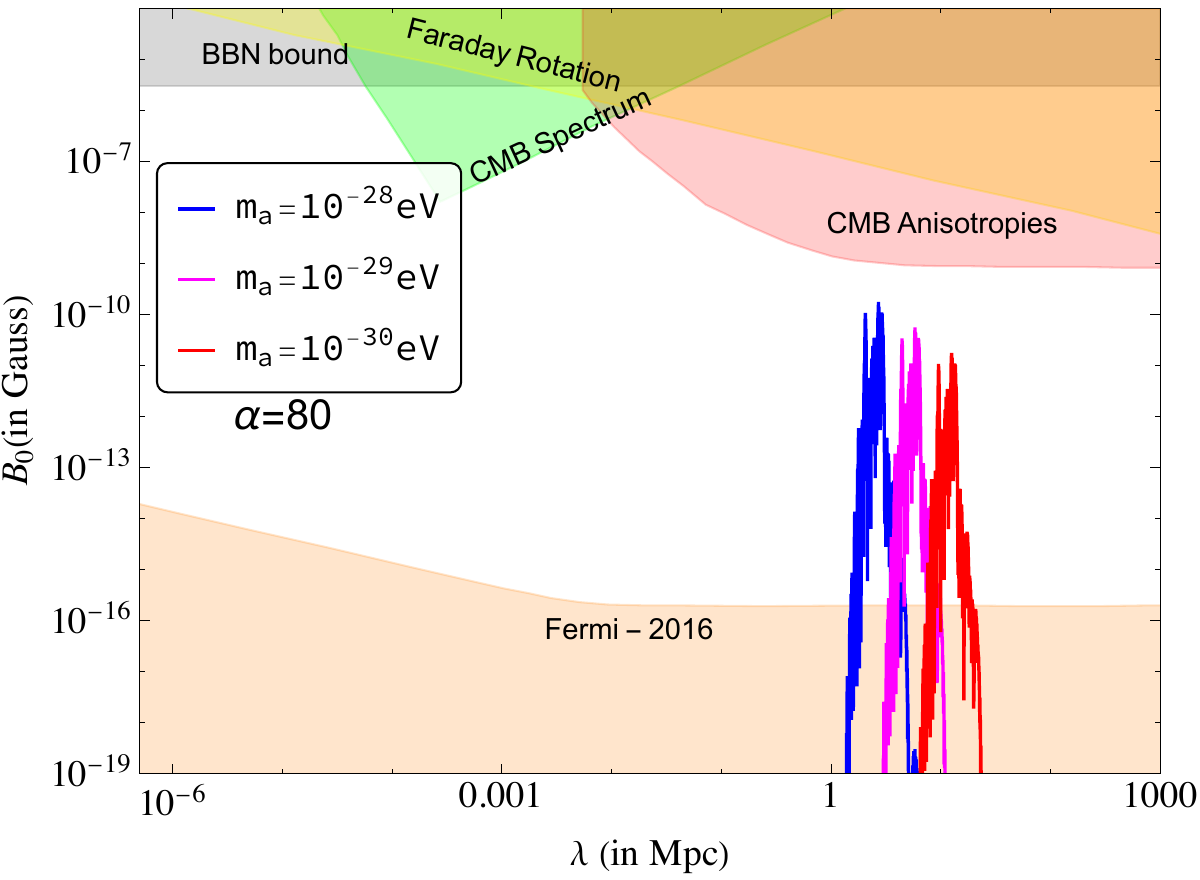}
\caption{In the figure, we have plotted the present-day magnetic field strength (in Gauss) as a function of the length scale (\(\lambda\)) along with different observational bounds. Here, we consider a fixed coupling parameter $\alpha=80$ with different colors indicating three different masses of the axion field.
}
\label{b0_bound}
\end{figure}

In Fig.~(\ref{b0_bound}), we plot the present-day magnetic field strength as a function of wavelength for the same coupling constant \( \alpha = 80 \), using the same three values of the ALP mass. Since these relatively light ALPs begin oscillating after photon decoupling, they preferentially amplify gauge fields on the corresponding length scales, as shown in the figure. We also include current observational bounds on large-scale magnetic fields for comparison. From the plot, it is clear that for coupling values \( \alpha > 70 \), the ALPs can boost magnetic field strengths on Mpc scales to levels consistent with all existing constraints. Specifically, for \( \alpha = 80 \), the maximum field strength achieved at around 1 Mpc is \( B_0(\sim 1\,\mathrm{Mpc}) \sim 10^{-10} \) Gauss. 
Although larger values of the coupling parameter \( \alpha > 80 \) can be considered, they tend to produce a significant amount of gravitational waves (GWs) at CMB scales. This may lead to a violation of current bounds on the tensor-to-scalar ratio. The final section will discuss the gravitational wave spectrum in detail.

\paragraph{\underline{Computing the Relic Density of the Axion:}\\}
The initial energy density of the axion can be approximated as  
\begin{align}
    \rho_\mfa(T_{\rm osc}) \simeq \frac{1}{2} \ma^2 \fa^2 \theta_i^2,
\end{align}
where \(\fa \) is the decay constant of the axion, and \( \theta_i \) is the initial misalignment angle.

The energy density redshifts to the present day due to the expansion of the background universe. Axions begin to oscillate when the background temperature equals \( T_{\rm osc} \), i.e., \( T \simeq T_{\rm osc} \). For \( T < T_{\rm osc} \), the axion behaves like non-relativistic matter, and its energy density dilutes as \( \rho_\chi \propto a^{-3} \). 

Assuming that no significant entropy production occurs after reheating, which is a reasonable approximation of the conservation of entropy, \( s(T)a^3 = \text{const} \), allows us to express the redshift factor as
\begin{align}\label{eq:a_a0}
    \frac{\aosc}{a_0}=\frac{T_0}{T_{\rm osc}}\l(\frac{g_{*}(T_0)}{g_*(T_{\rm osc})}\r)^{1/3}
\end{align}
where $g_*(T)$ is the effective number of entropy degrees of freedom at temperature $T$.

The relic density parameter $\Omega_\mfa h^2$ is define as $\Omega_\mfa h^2=\rho_\mfa(T_0)/\rho_{\rm cri}(T_0)$.  Now utilizing Eq.\eqref{eq:a_a0}, we can find that
\begin{align} \label{axiobaban}
    \Omega_\mfa h^2 \simeq 0.12 ~ \theta_i^2 \l(\frac{\fa}{10^{12}~\text{GeV}}\r)^2 \l( \frac{\ma}{10^{-5}~\text{eV}}\r)^{1/2}
\end{align}
As previously discussed, at early times when \( H \gg \ma \), the axion field remains in the slow-roll regime with \(\dot{\mfa} \approx 0\). In this phase, the equation of state of the axion is given by \( w_\mfa = P_\mfa / \rho_\mfa \approx -1 \), indicating that the field behaves as a dark energy component. Here, \(\rho_\mfa\) and \(P_\mfa\) represent the energy density and pressure of the background axion field, respectively.
At later times, when $ H \ll \ma $, the field undergoes coherent oscillations around the potential minimum, leading to an equation of state that oscillates around zero. Consequently, the energy density follows the scaling relation $\rho_\mfa \propto a^{-3}$, which is characteristic of non-relativistic matter. This implies that after the onset of oscillations, the axion behaves as a dark matter-like component. However, note that in the axion mass range of our interest $ \ma\leq 10^{-28}~\mathrm{eV}$, the onset of oscillations $\aosc > \aeq$ occurs much later compared to the matter-radiation equality. Therefore, axions cannot constitute the entirety of the dark matter component of the universe. Observational constraints from galaxy surveys place an upper bound on the relic density of axions, $\Omega_\mfa h^2 < 0.004$, for axion masses in the range $10^{-31}~\mathrm{eV} < \ma < 10^{-28}~\mathrm{eV} $ at 95\% confidence level~\cite{Lague:2021frh}. Utilizing the abundance equation for the axion Eq.\eqref{axiobaban}, one obtains a constraint on the axion decay constant,
\begin{equation}
\fa\lesssim 9.42\times10^{16}\,(10^{-28}\eV/\ma)^{1/4}\,\Gev .
\end{equation}
Here we assume the initial angle of misalignment $\theta_i \simeq 1$. Specifically, for ultralight axions in this mass range, we find that the maximum allowed decay constant is $\fa < 9.42 \times 10^{16}~\mathrm{GeV}$ for $\ma = 10^{-28}~\mathrm{eV}$, and that is sub-Planckian.

In our analysis, we restrict ourselves to values of the coupling constant \( \alpha < 100 \) to ensure the generation of a sufficient primordial magnetic field strength. This choice of \( \alpha \) and \( f_a \) is also consistent with the current upper limit on the axion-photon coupling, typically denoted as \( g_{a\gamma} = \alpha / f_a \ll 5.8 \times 10^{-11}~\text{GeV}^{-1} \)~\cite{PhysRevLett.133.221005}. Moreover, this selection remains in agreement with current observational constraints on large-scale magnetic fields~\cite{Neronov:2010gir, Tavecchio:2010mk, Dolag:2010ni, Dermer:2010mm, Vovk:2011aa, Taylor:2011bn, Takahashi:2011ac}.

\section{Secondary Gravitational Waves due to Magnetic Fields}\label{sec_IV}
The presence of primordial magnetic fields in the early universe can serve as a crucial source of gravitational waves. Metric perturbations induce anisotropic stress associated with helical or non-helical magnetic fields, thereby generating a stochastic gravitational wave background (SGWB). This mechanism provides unique insights into the dynamics of axions and gauge fields.  

As discussed, in the presence of ultralight axion-like particles (ALPs), the gauge field can be excited due to the oscillations of ALPs via a \( \mfa F \tilde{F} \) coupling. This excitation can significantly contribute to the production of gravitational waves.

Considering the tensor fluctuations, the FLRW metric can be written as
\begin{align}
    ds^2=a^2(\eta)\l[ -d\eta^2+(\delta_{ij}+h_{ij})dx^idx^j\r]
\end{align}
The tensor mode $h_{ij}$ is the traceless tensor, i.e., $\partial^i h_{ij} = h^i_i = 0$. In Fourier space, the equation of motion for the gravitational wave amplitude `$h$', for either polarization $h_{+}$ or $h_{-}$, becomes \cite{PhysRevD.65.023517, Sorbo:2011rz, Caprini:2014mja, Ito:2016fqp, Sharma:2019jtb, Okano:2020uyr}:
\begin{align}\label{eq:hk_S}
    \hk'' + 2\mH \hk' + k^2 \hk = \mS_\lambda(\vk, \eta),
\end{align}
where $\mS_{\lambda}$ is the source term. Here, we consider electromagnetic fields as the source term, which can be written as \cite{PhysRevD.65.023517, Sorbo:2011rz, Caprini:2014mja, Ito:2016fqp, Sharma:2019jtb, Okano:2020uyr, Maiti:2024nhv}
\begin{align}
    \mS_{\lambda}(\vk, \eta) = -\frac{1}{a^2(\eta)} \int \frac{d^3q}{(2\pi)^{3/2}} e^{ij}_{\lambda}(\vk) \left\{ A'_i(\vq, \eta) A'_j(|\vk - \vq|, \eta) - \epsilon^{iab} q_a A_b(\vq, \eta) \epsilon^{icd} (k - q)_c A_d(|\vk - \vq|, \eta) + \dots \right\}.
\end{align}
Here $e^{ij}_{\lambda}(\vk)$ is the polarization tensor corresponding to the mode with wave vector `$\vk$' and the polarization index $\lambda$. 
$\epsilon^{iab}$ is the three dimensional flat space 
Levi-civita symbol. As our source is helical so we chose circular polarization basis which can be define as $e^{ij}_{\lambda}=-\frac{1}{2}\l(\hat{e}^1\pm i~\hat{e}^2\r)_i\times\l(\hat{e}^1\pm i~\hat{e}^2\r)_j$,~\cite{Sharma:2019jtb}. Here $\hat{e}^{1}\,\&\, \hat{e}^{2}$ are a set of mutually orthonormal basis vectors of our coordinate system. Whereas the tensor power spectrum $\Pt(k,\eta)$ is define as $\Pt(k,\eta)=\frac{k^3}{2\pi^2}\l|\hk(k,\eta)\r|^2$~\cite{PhysRevD.65.023517, Sorbo:2011rz, Caprini:2014mja, Ito:2016fqp, Sharma:2019jtb, Okano:2020uyr, Maiti:2024nhv, Chakraborty:2024rgl}. It is convenient to express the tensor power spectrum as the sum of two components: one arising from vacuum fluctuations and the other from source terms. These two contributions are uncorrelated, so we write them as:
\begin{align}\label{powe_spl}
\Pt(k, \eta) = \Ptp(k, \eta) + \Pts(k, \eta),
\end{align}
where $\Ptp(k, \eta)$ refers to vacuum production, and $\Pts(k, \eta)$ is due to the EM field. The primary one is produced due to quantum fluctuations, which are well-known and near CMB scale; the spectrum remains scale invariant (for instance, see~\cite{1979JETPL..30..682S, Grishchuk:1974ny, Guzzetti:2016mkm, Maiti:2024nhv}). Here, we mainly focus on the secondary productions of the GWs due to late time dynamics, and we can write the tensor power spectrum induced by the EM field in terms of the electromagnetic power spectrum as~\cite{Maiti:2024nhv}
\begin{align}\label{pt-gen}
\Pts(k,\eta_0) =\frac{2}{\Mp^4} \int_{0}^{\infty}\frac{dq}{q}&\int_{-1}^1d\mu \frac{f(\mu,\beta,\lambda)}{[1+(q/k)^2-2\mu(q/k)]^{3/2}}\nn\\
&\times\l[\int_{\eosc}^{\eta_f}d\eta_1a^2(\eta_1)\Gk(\eta_f,\eta_1) {\Pb}^{1/2}(q,\eta_1){\Pb}^{1/2}(|\textbf{k}-\textbf{q}|,\eta_1)\r]^2 .
\end{align}
The resonant production of the gauge field due to the oscillating axion is typically narrowly peaked both in time and momentum. Therefore, we perform the above integral
Eq.\eqref{eq:hk_S} within $\eosc$ and $\eta_f$
from the beginning and end of gauge field production. Beyond this point, the magnetic field energy density dilutes as \( a^{-4} \) due to the expansion of the background universe.
Here $\Gk(\eta,\eta_1)$ is the Green's function associated with Eq.\eqref{eq:hk_S}, and   
the Green's function in the matter-dominated phase is written as \cite{Maiti:2024nhv}
\begin{align}
    \Gk^{\rm ra}(k,\eta,\eta_1)=\Theta(\eta-\eta_1)\frac{1}{k}\frac{x_1}{x^3}\l\{ -(x-x_1)\cos(x-x_1)+(1+xx_1)\sin(x-x_1)\r\},
\end{align}
where we have defined $x=k\eta$ is a dimensionless variable.

The tensor power spectrum is proportional to \( B_0^4(k) \) and can be expressed as  
\begin{align}
    \Pts(k,\eta_f)\propto\frac{B_0^4(k)\xeq^8}{\Mp^4k^4\aeq^4}~\mF(k)\times\overline{ \mI^2(k,\eosc,\eta_f)}
\end{align}
where $\overline{ \mI^2(k,\eosc,\eta_f)}$ carry the time integral part and $\mF(k)$ carry the momentum integral part of the above Eq.\eqref{pt-gen}. Here, the $\overline{ \mI^2(k,\eosc,\eta_f)}$ is the oscillation average of the time integral part. In the above, we have replaced the scale factor during the matter-dominated era as $a(\eta)=\aeq(\eta/\eeq)^2$. 

\begin{figure}[t]
\centering
\includegraphics[scale=0.41]{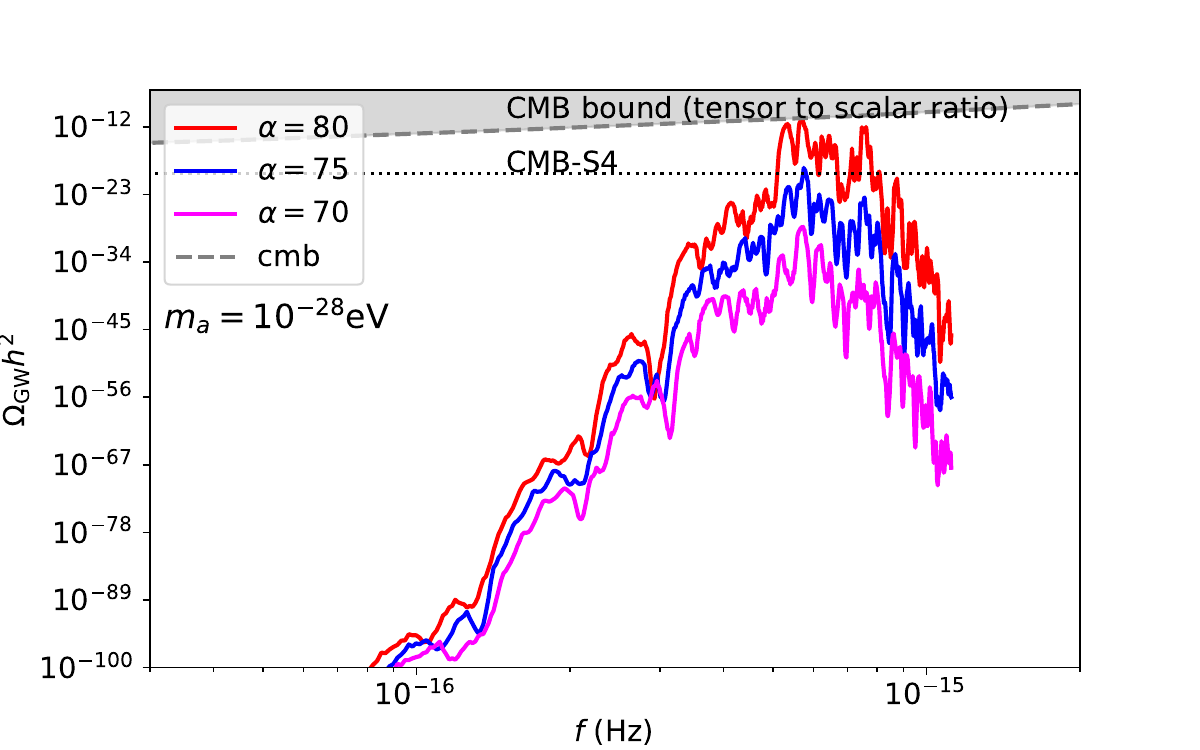}
\includegraphics[scale=0.41]{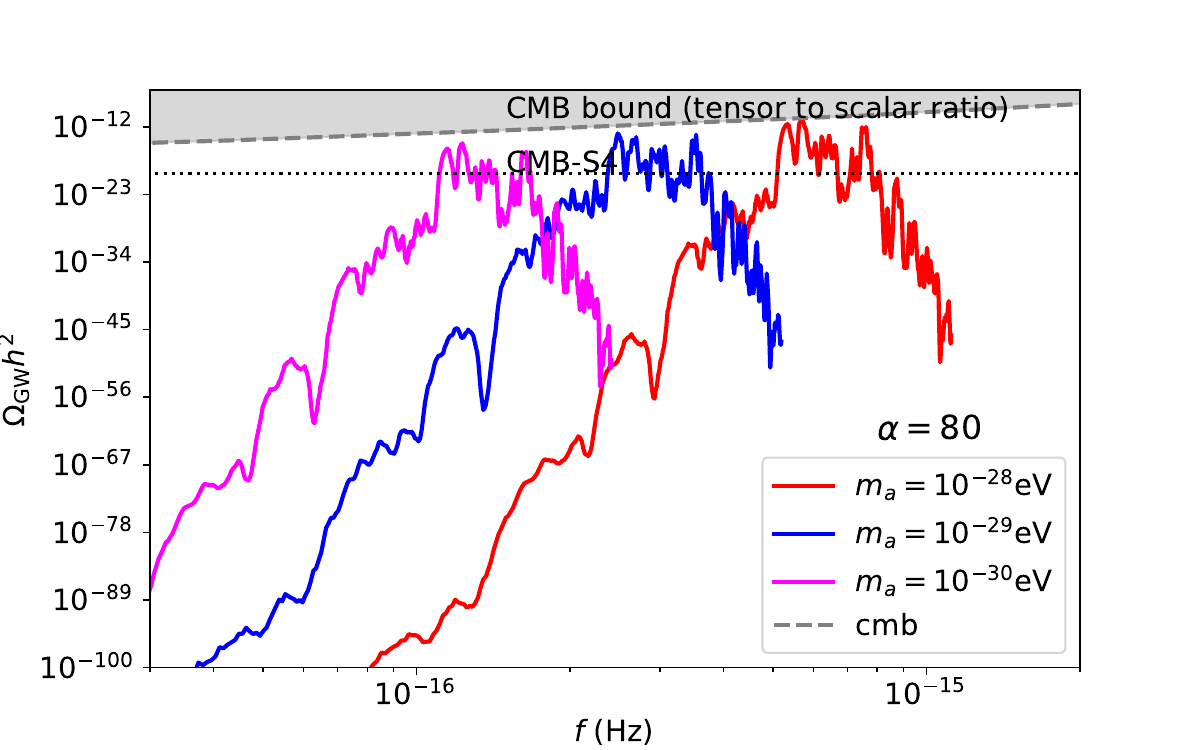}
\caption{In the figure above, we present the spectrum of stochastic gravitational waves (SGWs) induced by the magnetic field generated through the oscillation of axion-like particles (ALPs) at late times, after recombination. The quantity \( \ogwh \) is plotted as a function of the comoving frequency \( f \) (in Hz). The left panel illustrates the SGW spectrum for a fixed axion mass, \( \ma = 10^{-28}~\eV \), with different colors representing various values of the coupling parameter \( \alpha \). In the right panel, we explore the dependence of the SGW spectrum on the ALP mass while keeping the coupling parameter fixed at \( \alpha = 80 \). In both panels, the gray-shaded region denotes the CMB bound on the tensor-to-scalar ratio. The black dotted line in both figures is the future CMB-S4 observation of the tensor-to-scalar ratio corresponding to $r\simeq10^{-4}$.  For both spectra, we assume that the initial magnetic fields were generated during inflation due to a slow-roll-type curvature power spectrum, characterized by $A_s\simeq 2.1\times 10^{-9}$ and $\ns\simeq0.965$.
}
\label{gws}
\end{figure}

Now the dimensionless spectral energy density (SED) of secondary GWs \textit{today} (i.e., at $\eta_0$) is defined as $\ogw(k,\eta_0)=\rhogw(k,\eta_0)/3H_0^2\Mp^2$, where $H_0$ denotes the present value of the Hubble parameter. We can further express the  SED of SGWs in terms of radiation energy density as~\cite{Maiti:2024nhv}
\begin{align}
    \ogwh\simeq\Omega_{\rm ra}^0h^2~
    \ogw(k,\eta_f)
\end{align}
where $\Omega_{\rm ra}^0h^2\simeq4.3 \times 10^{-5}$, fractional radiation energy density at present time. 

In Fig.~(\ref{gws}), we present the present-day gravitational wave (GW) spectrum as a function of the observable frequency \( f \) (in Hz) for two different scenarios. 

In the left panel of Fig.~(\ref{gws}), we illustrate the dependence of the stochastic gravitational wave (SGW) spectrum on the coupling parameter \( \alpha \) for a fixed axion-like particle (ALP) mass, \( m_a = 10^{-28}~\eV \). We observe that as the coupling strength increases, the amplitude of the GW spectrum also increases, making it potentially detectable by future CMB experiments~\cite{CMB-S4:2020lpa}. However, the overall shape of the spectrum remains unchanged, and in all cases, the peak appears at the same position. This behavior arises because the peak position of the GW spectrum is determined by the magnetic field spectrum, which, in turn, depends on the ALP mass.  

In the right panel of Fig.~(\ref{gws}), we plot $\ogwh$ as a function of $f$ (in Hz), where different colors represent different values of the ALP mass, which oscillates after recombination. Here, we fix the coupling constant at $\alpha = 80$. We observe that varying the ALP mass leads to a significant shift in the peak position of the GW spectrum. This is expected, as a lower ALP mass implies a later onset of oscillations, and only those modes that re-enter the horizon during this period experience tachyonic instability. Additionally, we find that decreasing the ALP mass results in a gradual decrease in the peak amplitude of the GW spectrum for the same coupling $\alpha = 80$. This effect can be attributed to the initial spectral behavior of the magnetic field, which originates from inflationary curvature perturbations.
In this study, we have also found that the generated magnetic field exhibits a helical nature, which naturally implies the production of helical gravitational waves (GWs). This results in an asymmetry between the two polarization modes of the tensor power spectrum, leading to a distinctive signature in the B-mode polarization of the cosmic microwave background (CMB). If future CMB experiments, such as CMB-S4~\cite{CMB-S4:2020lpa}, detect any helical features, it will serve as a clear indication of the existence of a helical magnetic field. Moreover, such a detection could indirectly probe the presence of ultralight axion-like particles.

Furthermore, we find that consistency with the tensor-to-scalar ratio constraints from Planck 2018~\cite{Planck:2018vyg} imposes a strong bound on the coupling parameter $\alpha$. Specifically, we observe that $\alpha = 80$ leads to significant GW production at CMB scales, whereas values exceeding $\alpha > 80$ would immediately violate the current observational constraints on the tensor-to-scalar ratio.

\section{Conclusion}\label{sec_V}
Observations across different astrophysical and cosmological scales indicate that our universe is fully magnetized. However, explaining the origin of these cosmic magnetic fields remains a significant challenge. A natural mechanism for generating such fields is highly desirable, yet existing models often face issues such as strong coupling or backreaction. Even in cases where these issues are circumvented, the generation of large-scale magnetic fields typically requires non-trivial couplings with the gauge field, whose origins remain elusive.

In this work, we address this issue by investigating the generation of magnetic fields across all relevant length scales from curvature perturbations. These perturbations naturally arise from quantum fluctuations during inflation. In the presence of curvature, the spacetime metric exhibits deviations from perfect spatial flatness, and these metric fluctuations can break the conformal flatness property of the background, leading to non-trivial gauge field production at higher orders in perturbation theory. Exploiting this breaking, we first derive how curvature-induced metric fluctuations induce gauge field production and determine their impact on the resulting magnetic field spectrum.  As expected, the resulting magnetic field strength remains small due to the relatively small amplitude, as it was generated gravitationally. 
Interestingly, we find that a primordial black hole (PBH)-type curvature power spectrum can enhance the magnetic field strength by up to a fourth order of magnitude compared to the case with a simple slow-roll type curvature power spectrum.
For example, we consider a scale non-invariant power spectrum with a peak frequency $k_p$, typically associated with the formation of primordial black holes (PBHs). For wavenumbers \( k < k_p \), the magnetic field strength becomes sensitive to the peak frequency of the curvature perturbations, resulting in an amplification of the magnetic field by up to four orders of magnitude, even at Cosmic Microwave Background (CMB) scales.
We found present day strength of the magnetic field follows $B_0 \propto\sqrt{ A_0 k_p} k^\frac32$ for $k < k_p$, and 
        $\sqrt{ A_s} k^\frac32 $ for $k > k_p$. 
For a PBH-type spectrum, if we assume $A_0 \simeq 1$, the maximum magnetic field strength on the \(1\,\Mpc\) scale is found to be of the order \(B_0 \sim 10^{-41}\,\mathrm{G}\), whereas for a slow-roll type curvature spectrum, the corresponding strength is approximately \(B_0 \sim 10^{-45}\,\mathrm{G}\). 
Moreover, a distinct scaling behavior emerges in the magnetic field spectrum, which follows a power-law dependence \(B(k)\propto k^{3/2}\). The total energy density associated with the generated magnetic field is found to be substantial and may have a significant impact on the evolution of the early universe.

Additionally, we explore the effects of the post-inflationary reheating phase on the evolution of gauge fields. Our analysis reveals that non-trivial reheating scenarios can significantly enhance gauge field production. In scenarios where the reheating equation-of-state parameter satisfies $\wre>1/3$, we find that the magnetic field can carry a significant fraction of the total energy density. This enhancement has the potential to influence the standard predictions of Big Bang Nucleosynthesis (BBN). In particular, we demonstrate that in kination-like reheating scenarios with $\wre=1.0$ and a background temperature of  $T \sim 10^8\,\Gev$ , the energy density of the magnetic field can surpass that of the inflaton. Such a situation could lead to a magnetically dominated universe during the post-inflationary epoch, thereby altering the dynamics of BBN and the subsequent thermal history of the universe.

As previously discussed, the gravitationally generated magnetic fields sourced by inflationary curvature perturbations are typically too weak on large scales $(\sim\Mpc)$ to meet current observational bounds. Therefore, additional dynamics are required to further amplify these seed magnetic fields at later times. In this context, we explore the possibility of axion-photon coupling as a mechanism for such enhancement.
Axions naturally couple to gauge fields through the Chern-Simons term $\frac{\alpha\,\mfa}{\fa}F\tilde{F}$, where $\alpha$ is the dimensionless coupling constant and $\fa$ is the axion decay constant. When the axion field begins to oscillate, which occurs when the Hubble parameter satisfies $ 3H \simeq \ma$ (with $\ma$ denoting the axion mass), a tachyonic instability is induced in the gauge field sector. We find that for coupling constants in the range \( 60 < \alpha < 80 \) and axion masses $ 10^{-31} ~\mbox{eV} \lesssim~\ma \lesssim 10^{-28}\,\eV$, the gauge fields originally generated by inflationary curvature perturbations can be significantly amplified. As a result, the magnetic field strength can reach values as high as \( B_0 \sim 10^{-10}~\mathrm{G} \) on Mpc scales.

In the final section, we analyze the generation of secondary gravitational waves (SGWs) induced by magnetic fields in the presence of axions. Our results indicate a substantial production of SGWs, with amplitudes potentially within the sensitivity reach of future Cosmic Microwave Background (CMB) experiments, such as CMB-S4. To remain consistent with current observational bounds on the tensor-to-scalar ratio, we find that the axion-photon coupling strength must satisfy \( \alpha \leq 80 \).
An intriguing feature of this mechanism is the natural generation of helical magnetic fields, which in turn lead to the production of helical gravitational waves. The future detection of helical B-mode polarization in the CMB would provide compelling evidence for the presence of such helical magnetic fields, thereby lending support to the proposed amplification mechanism.
Moreover, we observe that the axion mass plays a crucial role in determining the peak frequency of the resulting gravitational wave spectrum. This mass-dependent feature imprints a distinctive signature on the CMB power spectrum, offering a promising observational channel to probe the existence of ultra-light axions through precision CMB measurements.

Note added: While working on our present paper, we came across a recent paper \cite{brandenberger2025} where the generation of large-scale magnetic fields after recombination has been discussed. 

\acknowledgments
SM wishes to thank the Council of Scientific 
and Industrial Research, Ministry of Science and Technology, Government 
of India (GoI), for financial assistance.
DM gratefully acknowledge the support received from the Science and
Engineering Research Board, Department of Science and Technology, GoI, through 
the Core Research Grant~CRG/2020/003664.
We wish to thank the Gravitation and High Energy Physics Groups at IIT Guwahati 
for illuminating discussions.

\appendix

\section{Detailed calculation of gauge field generation in inflationary perturbed background}
\label{appendix-A}

We first begin with the action of the background Lagrangian and then perturb the fields and the metric, 
\begin{equation}
    S = \int  d^4x \ \sqrt{-g} \left[ \frac{M^2_{pl}}{2} R +\frac{1}{2} \partial^\mu \varphi \partial_\mu \varphi - V(\varphi) - \frac{1}{4} F_{\mu \nu} F^{\mu \nu} \right].
\end{equation}
Now, finding the equation of motion for the gauge field, we arrive at,
\begin{equation}
\partial_\nu \left( \frac{\delta \mathcal{L}_{EM}}{\partial \partial_\mu A_\nu } \right)= \partial_\nu \left( \frac{\partial \sqrt{-g} F_{\mu\nu} F^{\mu\nu}}{\partial \partial_\mu A_\nu} \right) = \frac{\partial}{\partial x^\mu} [\sqrt{-g} g^{\alpha \mu}g^{\beta \nu} F_{\alpha \beta}] = 0,
\end{equation}
Expressing the gauge field in terms of Fourier modes,
\begin{equation}\label{gauge}
    A_i = \int \frac{d^4 x}{(2\pi)^{3/2} }\epsilon_i^{\lambda} A^{\lambda}_{\veck} e^{i(\veck\cdot {\bf x} - k \eta)}
\end{equation}
The equation of motion of the gauge field mode in the presence of the metric fluctuations in Fourier space is governed by
\begin{align}
    A_{\vk}^{\lambda ''}+ A_{\vk}^{\lambda}=\frac{1}{k^2}\mathcal{J}_{\vk}^{\lambda}(x),
\end{align}
Here, the prime $(')$ is defined as the derivative with respect to the new dimensionless variable $x=k\eta$, and $\Jik(\eta)$ is the source due to metric fluctuations. \footnote{We use the following convention $\Tilde{F}(\Vec{q},\eta) = \frac{1}{(2\pi)^{3/2}} \int d^3x \ e^{i\Vec{q}\Vec{x}} F(\Vec{x},\eta)$.}
Now, the solution of the homogeneous part of the above Eq.(\ref{gauge}) is
\begin{align}
    A_{\vk}^{\lambda}=C_1 \cos(k\eta)+C_2\sin(k\eta).
\end{align}
Considering two independent solutions, we can contract the Green's function,
\begin{align}
    \mathrm{G}_k(x,x_1) &=\Theta(x-x_1) \frac{A_{\vk,1}^{\lambda}(x)A_{\vk,2}^{\lambda}(x_1)-A_{\vk,1}^{\lambda}(x_1)A_{\vk,2}^{\lambda}(x)}{A_{\vk,1}^{\lambda '}(x_1)A_{\vk,2}^{\lambda}(x_1)-A_{\vk,1}^{\lambda}(x_1)A_{\vk,2}^{\lambda '}(x_1)}\\
   & =\Theta\frac{A_{\vk,1}^{\lambda}(x)A_{\vk,2}^{\lambda}(x_1)-A_{\vk,1}^{\lambda}(x_1)A_{\vk,2}^{\lambda}(x)}{-C_1C_2}
   =\Theta(x-x_1)\sin(x-x_1).
\end{align}
Now the solution of the above equations is,
\begin{align}
    A^{\lambda}_{\vk}=A_{\vk}^{\rm vac}+ \frac{1}{k^2}\int dx_1 G_{k}(x,x_1)\mathcal{J}_{\vk}^{\lambda}(x_1)
    =A_{\vk}^{\rm vac}+ \frac{1}{k} \int d\eta_1 \sin(k(\eta-\eta_1) \Jik(\eta_1) .
\end{align}
We are assuming that metric perturbations vanish before inflation starts so that we can define an appropriate initial conformal vacuum state. Because of the presence of the inhomogeneous background in the asymptotic future, this solution will behave as a linear superposition of positive and negative frequency modes with different momentum and different polarizations, i.e.
\begin{align}
    A_{k}^{\lambda}(\eta\rightarrow\infty) \sum_{\lambda'}\sum_{q} \l( \alpha_{\vk}^{\lambda\lambda'}(q,\eta) \frac{\epsilon_{\vq}^{\lambda}}{\sqrt{2q}} e^{i(\vq\cdot\vx-q\eta)} + \beta_{\vk}^{\lambda\lambda'}(\vq,\eta) \frac{\epsilon^{\lambda *}_{\vq}}{\sqrt{2q}}e^{-i(\vq\cdot\vx -q\eta)} \r),
\end{align}
here $\alpha_{\vk}^{\lambda\lambda'}~\&~\beta_{\vk}^{\lambda\lambda'}$ are the Bogoliubov coefficient to the first order in the metric perturbations. Now comparing this expression with Eq.(\ref{gauge}), it is straight forward to obtain the Bogoliubov coefficient $\beta_{\vk}^{\lambda\lambda'}(q,\eta)$, which 
is given by,
\begin{align}
    \beta_{\vk}^{\lambda\lambda'}(q,\eta)=-\frac{i}{\sqrt{2k}}\sum_i \int_{\eta_i}^{\eta}\epsilon_{i\vq}^{\lambda *}\Jik(\vk+\vq,\eta_1) e^{-ik\eta_1}d\eta_1 .
\end{align}
Now the total number of photon created with comoving wave number '$k/a$' is defined as
\begin{align}
    N_{\vk}(\eta)=\sum_{\lambda,\lambda'}\sum_{q}\l| \beta_{\veck}^{\lambda\lambda'}(q,\eta)\r|^2=\int_{k_{*}}^{\ke} \frac{d^3q}{(2\pi)^{3}}\l| \beta_{\veck}^{\lambda\lambda'}\r|^2 
    =\int_{k_{*}}^k \frac{d^3q}{(2\pi)^{3}} \l| \beta_{\veck}^{\lambda\lambda'}\r|^2 +\int_k^{\ke}\frac{d^3q}{(2\pi)^{3}}\l| \beta_{\veck}^{\lambda\lambda'}\r|^2.\label{number}
\end{align}
The electromagnetic source term $\Jik$ is sourced by the metric fluctuations $\Phi$ and it can be written as \cite{Maroto:2000zu}
\begin{align}
   \Jik=-\sqrt{2q}\l[ \l( i\Phi'(\vk+\vq,\eta)+\frac{q^2-\veck\cdot\vecq}{q}\Phi(\veck+\vecq,\eta)\r)\epsilon_{i\vecq}^{\lambda} e^{-iq\eta} +( \epsilon_{\vecq}^{\lambda}\cdot\veck) \Phi(\veck+\vecq,\eta) \frac{\vq_i}{q} e^{-iq\eta}\r.\nn\\
   \l.\underbrace{ -i\frac{\epsilon_{\vecq}^{\lambda}\cdot\veck}{q^2}\frac{\partial}{\partial \eta}\l[ \Phi(\veck+\vecq,\eta)e^{-ik\eta} \r]\veck_i} \r] .
\end{align}
Last term of the above equation does not contribute to the $\beta_{\vk}^{\lambda\lambda'}$ because of the transverse condition of the polarization vectors. So for the further calculations we drop this term, multiplying this with the polarization vector and taking the modulus square, we get
\begin{align}
    \l|\epsilon_{\vk}^{\lambda}\cdot J^{\veck,\lambda'}\r|^2=2q\l[ |\Phi'(\vk+\vq,\eta)|^2 + \l( \frac{q^2-\vk\cdot\vq}{q}\r)^2 \Phi^2(\vk+\vq,\eta)(\epsilon_{\vq}^{\lambda'}\cdot\epsilon_{\vk}^{\lambda})^2  + \frac{( \epsilon_{\vq}^{\lambda'}\cdot\vk)^2(\epsilon_{\vk}^{\lambda}\cdot\vq)^2}{q^2}  \Phi^2(\vk+\vq,\eta)\r],\\
    =\frac{4\pi^2q}{|\vk+\vq|^3}\l[ \l( \mPphip(\vk+\vq,\eta) + \l( \frac{q^2-\vk\cdot\vq}{q}\r)^2\mPphi(\vk+\vq,\eta) \r)(\epsilon_{\vq}^{\lambda'}\cdot\epsilon_{\vk}^{\lambda})^2 +  \frac{( \epsilon_{\vq}^{\lambda'}\cdot\vk)^2(\epsilon_{\vk}^{\lambda}\cdot\vq)^2}{q^2} \mPphi(\vk+\vq,\eta) \r],
\end{align}
here $\mPphi~\&~\mPphip$ are defined as
\begin{align}
   \mPphi(\vk+\vq,\eta)=\mathcal{P}_{\Phi}^i(|\vk+\vq|)\mC^2(|\vk+\vq)\eta);\, \mPphip(\vk+\vq,\eta)==\mathcal{P}_{\Phi}^i(|\vk+\vq|)\mC^{'2}(|\vk+\vq)\eta).
\end{align}
In the above integral we kept only those therm which are contributing in the number spectrum. Now we can write the number spectrum as
\begin{align}
    N_k=\sum_{\lambda\lambda'}\int \frac{d^3q}{(2\pi)^{3}}\l(\frac{q}{k}\r)\Phi_0^2(|\veck+\vecq|) {\cal D}(k,q), 
    \end{align}
    where we define
    \begin{align}
 {\cal D}(k,q) =  \l|\int d\eta' \l\{ \l( i \mC^{'}(|\vk+\vq|\eta') + \frac{q^2-\veck\cdot\vecq}{q}\mC(|\vk+\vq|\eta') \r)(\epsilon^{\lambda}_{\veck}\cdot\epsilon^{\lambda'}_{\vecq}) +
    \frac{(\epsilon^{\lambda}_{\veck}\cdot\vecq)(\epsilon^{\lambda'}_{\vecq}\cdot\veck)}{q^2}\mC(|\vk+\vq|\eta')  \r\}\r|^2   .
\end{align}

\section{Computing the Number Spectra}

\subsection{Computing the total number of photons for the comoving wavenumber $k$ induced by the metric fluctuations:}
In this section, we discuss the gravitational production of gauge fields sourced by metric fluctuations that originate from quantum fluctuations of the inflaton field during inflation. These metric perturbations, although frozen on super-horizon scales during inflation, can source gauge field production after inflation as they re-enter the horizon.

The total number of photons produced for a given comoving wavenumber \( k \) due to these post-inflationary metric fluctuations can be expressed as
\begin{align}
    N_k=\sum_{\lambda\lambda'}\int \frac{\d q^3}{(2\pi)^3}\frac{q}{k}\l|\int_{\eta_k}^{\ee}d\eta'\l\{\l(i\Phi'(|\vk+\vq|\eta')+\frac{q^2-\vk\cdot\vq}{q}\Phi(|\vk+\vq|\eta')\r)(\epsilon^{\lambda}_{\veck}\cdot\epsilon^{\lambda'}_{\vecq})+\frac{(\epsilon^{\lambda}_{\veck}\cdot\vecq)(\epsilon^{\lambda'}_{\vecq}\cdot\veck)}{q}\Phi(|\vk+\vq|\eta')\r\} \r|^2\label{eq:nk_i}
\end{align}
Here, \( \Phi(k\,\eta) \) denotes the metric perturbation generated by quantum fluctuations of the inflaton field during inflation. During the inflationary phase, the evolution of the metric perturbation \( \Phi(k,\eta) \) is governed by the energy scale of inflation and the first slow-roll parameter \( \epsilon \). The solution for \( \Phi \) can be approximated as \cite{Mukhanov:2005sc}
\begin{align}
    \Phi \simeq \frac{\dot{\phi}}{2\Mp^2} \frac{1}{\sqrt{2k^3}} \simeq \frac{\sqrt{\epsilon}}{2\sqrt{k^3}} \frac{H}{\Mp} \simeq \epsilon\,\mathcal{R},
\end{align}
where \( \dot{\phi} \) is the time derivative of the background inflaton field, which is related to the first slow-roll parameter by \( \dot{\phi} = \sqrt{2\epsilon}\,\Mp\,\HI \), with \( \HI \) denoting the Hubble parameter during inflation, assumed to be approximately constant. The time integral appearing in Eq.~\eqref{eq:nk_i} runs from \( \eta_k \), the conformal time at which the mode \( k \) exits the horizon, to \( \eta_e \), the conformal time at the end of inflation.
Alternatively, noting that 
\begin{align}
    \delta\varphi_k &= A_k \frac{V_{,\varphi}}{V}, 
    \Phi_k = 4\pi A_k \frac{\dot{\varphi}_0}{H} \frac{V_{,\varphi}}{V} = -\frac{1}{2} A_k \left( \frac{V_{,\varphi}}{V} \right)^2.
\end{align}
We also know that the comoving curvature perturbation is given by:
\begin{align}
    \mathcal{R}_k = A_k \frac{H}{\dot{\varphi}_0} \frac{V_{,\varphi}}{V}.
\end{align}
From the slow-roll equation of motion
\begin{align}
    3H \dot{\varphi}_0 \simeq -V_{,\varphi} \quad \Rightarrow \quad
    \dot{\varphi}_0 \simeq -\frac{V_{,\varphi}}{3H}.
\end{align}
Substituting into the expression for \( \Phi_k / \mathcal{R}_k \)
\begin{align}
    \frac{\Phi_k}{\mathcal{R}_k}
    &= -\frac{1}{2} \cdot \left( -\frac{V_{,\varphi}}{3H^2} \right) \cdot \frac{V_{,\varphi}}{V}
    = \frac{1}{6H^2} \left( \frac{V_{,\varphi}}{V} \right)^2.
\end{align}
Finally, remembering that in the slow-roll approximation, the Friedmann equation gives:
\begin{align}
    \frac{\Phi_k}{\mathcal{R}_k}
    &\simeq \frac{1}{6} \cdot \frac{3}{V} \left( \frac{V_{,\varphi}}{V} \right)^2
    = \frac{1}{2} \left( \frac{V_{,\varphi}}{V} \right)^2 \equiv \epsilon,
\end{align}
where \( \epsilon \) is the slow-roll parameter:
Since the power spectrum is proportional to the square of the mode amplitude
\begin{align}
    \mathcal{P}_\Phi(k) = |\Phi_k|^2 \simeq \epsilon^2 |\mathcal{R}_k|^2 = \epsilon^2 \mathcal{P}_{\mathcal{R}}(k) \label{eq:Prphi_a}.
\end{align}
For simplicity, we define the power spectrum of the metric perturbation \( P_{\Phi}(k) \) in terms of the comoving curvature power spectrum \( P_{\mathcal{R}}(k) \). During inflation, these two quantities are related by
\begin{align}
    P_{\Phi}(k) \simeq \epsilon^2 P_{\mathcal{R}}(k),
\end{align}
where \( \epsilon \) is the first slow-roll parameter. This relation holds only during the inflationary epoch. After inflation, the evolution of the metric perturbations follows a different relation, as described in Eq.~\eqref{eq:P_phi_post_i}. Using this expression, the total number of photons produced per comoving wavenumber \( k \) due to gravitational interactions with the metric perturbations can be written as
\begin{align}
    N_k=\frac{1}{2}\sum_{\lambda\lambda'}\int \frac{\d q}{q}\,\frac{q}{k}\frac{q^3}{2\pi^2}\int_{-1}^1\d \gamma\times\mD(k,q,\gamma)
\end{align}
where $\mD(k,u)$ is define as
\begin{align}
    \mD(k,q,\gamma) &=\l|\int_{\eta_i}^{\eta_f}\d \eta' \l\{\l(i\Phi'(|\vk+\vq|\eta')+\frac{q^2-\vk\cdot\vq}{q}\Phi(|\vk+\vq|\eta')\r)(\epsilon^{\lambda}_{\veck}\cdot\epsilon^{\lambda'}_{\vecq})+\frac{(\epsilon^{\lambda}_{\veck}\cdot\vecq)(\epsilon^{\lambda'}_{\vecq}\cdot\veck)}{q}\Phi(|\vk+\vq|\eta')\r\}  \r|^2\nn\\
   & =\l| \int_{x_i}^{x_f} \d x \l\{i\gamma\frac{\partial\Phi(\tu\,x)}{\partial x}+(1+u^2-2\gamma^2)\Phi(\tu\,x)\r\}\r|^2
\end{align}
where we define some new variable like $x=|k\eta|$ and $\tu=|1+\vq/\vk|=(1+u^2-2u\gamma)^{1/2}$. Here $\gamma=\hat{k}\cdot\hat{q}=\cos(\theta)$, where $\theta$ is the angle between $\vk$ and $\vq$.

During inflation, the evolution of the metric fluctuations is almost constant over time just after the corresponding modes leave the horizon. Now to compute $N_k$, we break the above integral into two limit $\int_{\kpv}^{\ke}\d q=\int_{\kpv}^k\d q+\int_k^{\ke}\d q$ and we can write $N_k$ as
\begin{align}
    N_k &\simeq\int_{\kpv}^k\frac{\d q}{q}\frac{q}{k}\frac{q^3}{2\pi^2}\epsilon^2\mR^2(k)\times\mathcal{O}(1)+\int_{k}^{\ke}\frac{\d q}{q}\frac{q}{k}\frac{q^3}{2\pi^2}\epsilon^2\mR^2(u\,k)\times\mathcal{O}(1)\\
    &\simeq \l\{\int_{\umin}^1\frac{\d u}{u}\,u^4\epsilon^2\mP_{\mR}(k)+\int_{1}^{\umax}\frac{\d u}{u}u^4\epsilon^2\frac{\mP_{\mR}(k\,u)}{u^3}\r\}\times\mathcal{O}(1)
\end{align}
Let us consider an ultra-slow-roll type inflationary scenario, where the enhanced curvature power spectrum is generated due to the ultra-slow-roll phase, and we consider that the comoving curvature power spectrum is parameterized as
\begin{align}
      \mathcal{P_R}(k) = A_s \left(\frac{k}{k_*}\right)^{n_s-1} + A_0~\text{Exp}\left[-\frac{(k-k_p)^2}{\delta\times k_p^2}\right],\label{eq:def_PR}
\end{align}
Here $A_s\simeq 2.1\times 10^{-9}$ is the amplitude of the scalar curvature power spectrum observed by Planck, and $\ns$ is the scalar spectral index~\cite{Planck:2018vyg}. Here, $A_0$ is the amplitude of the curvature power spectrum at peak wavenumber, and $\delta$ is the dimensionless parameter that sets the spectral shape of the curvature near the peak. $\kp$ is the position of the peak of the curvature power spectrum.

Now, if we consider only the contribution that came from the 1st part of the above Eq.\eqref{eq:def_PR}, we have found
\begin{align}
    N_k^{\rm v} &\simeq \frac{\epsilon^2A_s}{4}\l\{1-\l(\frac{\kpv}{k}\r)^4\r\}+\frac{\epsilon^2A_s}{\ns}\l(\frac{k}{\kpv}\r)^{\ns-1}\l\{ \l(\frac{\ke}{k}\r)^{\ns}-1\r\}\\
    &\simeq \epsilon^2 A_s\l\{ \frac{1}{4}+\frac{1}{\ns}\l(\frac{\ke}{\kpv}\r)^{\ns}\l(\frac{\kpv}{k}\r)\r\}\simeq \frac{\epsilon^2A_s}{\ns}\l(\frac{\ke}{\kpv}\r)^{\ns}\l(\frac{\kpv}{k}\r)\label{eq:nk_v}
\end{align}
Similarly, the contribution due to the PBH-type spectrum during inflation is
\begin{align}
    N_K^{\rm PBH} &\simeq \frac{\epsilon^2A_0}{4}\l\{ 1-\l(\frac{\kpv}{k}\r)^4\r\}+\sqrt{\pi\delta}A_0\epsilon^2\l(\frac{\kp}{k}\r)\l[1+\text{Erf}\l(\frac{\kp-k}{\kp\sqrt{\delta}}\r)\r ]\\
    &\simeq \epsilon^2 A_0\l\{ \frac{1}{4}+\sqrt{\pi\delta}\l(\frac{\kp}{k}\r)\l[1+\text{Erf}\l(\frac{\kp-k}{\kp\sqrt{\delta}}\r)\r]\r\}\simeq \epsilon^2A_0\sqrt{\pi\delta}\l(\frac{\kp}{k}\r)\l[1+\text{Erf}\l(\frac{\kp-k}{\sqrt{\delta}\kp}\r)\r]\label{eq:nk_i}
\end{align}
Now, if we combine Eq.\eqref{eq:nk_v} and Eq.\eqref{eq:nk_i}, we get the total photon number produced due to the metric fluctuations during inflation. SO the total number of photons produced at the end of inflation is
\begin{align}
    N_k\simeq \frac{\epsilon^2A_s}{\ns}\l(\frac{\kpv}{k}\r)\l\{ \l(\frac{\ke}{\kpv}\r)^{\ns}+\frac{A_0\ns\sqrt{\pi\delta}}{A_s}\l(\frac{\kp}{\kpv}\r)\l[ 1+\text{Erf}\l(\frac{\kp-k}{\kp\sqrt{\delta}}\r)\r]\r\}
\end{align}
As we see, the total number produced for co-moving wavenumber $k$ goes as $N_k\propto k^{-1}$. Now the energy density of the produced magnetic field per logarithmic wavenumber is defined as $\rhob(k,\eta)=(k/a(\eta))^4N_k(\eta)$. Or we can define the comoving magnetic energy density as $\tilde{\rhob}(k,\eta)=k^4N_k(\eta)$. Now, at the end of inflation, we can define the magnetic energy store per logarithmic $k$ as
\begin{align}
    \mP_{\rm B}(k,\ee)\simeq \frac{k^3\kpv}{a^4(\ee)}\frac{\epsilon^2A_s}{\ns}\l\{ \l(\frac{\ke}{\kpv}\r)^{\ns}+\frac{A_0\ns\sqrt{\pi\delta}}{A_s}\l(\frac{\kp}{\kpv}\r)\l[ 1+\text{Erf}\l(\frac{\kp-k}{\kp\sqrt{\delta}}\r)\r]\r\}\label{eq:mPbk_i}
\end{align}

\noindent
\subsection{Post-inflationary Productions}
The damping of gauge field modes due to conductivity arises from causal particle interactions, which can only influence sub-horizon modes, those within the causal horizon. In contrast, super-horizon modes, corresponding to spatial fluctuations on scales larger than the causal horizon, remain unaffected by such local processes, including magnetohydrodynamics (MHD) and conductivity effects. As a result, any mechanism that sources gauge field amplification on super-horizon scales will not experience suppression from conductivity.

In particular, curvature perturbations generated during inflation can continue to influence the evolution of the gauge field after inflation ends, sourcing its growth on super-horizon scales. This enhancement persists until the modes re-enter the horizon, at which point causal processes like conductivity become relevant. We demonstrate that the presence of a nontrivial curvature power spectrum can lead to significant gauge field production at late times, but only while the corresponding modes remain outside the horizon. Once the modes re-enter, the increasing conductivity efficiently damps further amplification. Moreover, we find that inside the horizon, the curvature perturbations decay rapidly with time and no longer contribute to gauge field excitation.

After a given mode exits the horizon during inflation, the corresponding curvature perturbation remains conserved on super-horizon scales. However, once inflation ends, these modes gradually re-enter the horizon and begin to decay over time (see Eq.~\eqref{eq:phi_sol}). During the post-inflationary era, the power spectra of the metric and curvature perturbations, $P_{\Phi}(k)$ and $P_{\mR}(k)$ respectively, are related via Eq.~\eqref{eq:P_phi_post_i}.
Importantly, these residual metric fluctuations can still amplify gauge fields after inflation, but only for modes that remain outside the horizon. Once a given mode re-enters the horizon, the high conductivity of the plasma causes the corresponding gauge field amplification to saturate and cease.

Since our primary interest lies in modes that are well inside the horizon today, we focus on computing the total number of photons produced due to these post-inflationary metric perturbations. This quantity can be expressed as

\begin{align}
    N_k=\sum_{\lambda,\lambda'}\int \frac{\d q^3}{(2\pi)^3}\frac{q}{k}\Phi_0^2(|\vk+\vq|)\mD(k,q)
\end{align}
where $\mD(k,q)$ is defined as~\cite{Maroto:2000zu}
\begin{align}
  \mD(k,q) =  \l|\int_{\ee}^{\eta_k} d\eta' \l\{ \l( i \mC^{'}(|\vk+\vq|\eta') + \frac{q^2-\veck\cdot\vecq}{q}\mC(|\vk+\vq|\eta') \r)(\epsilon^{\lambda}_{\veck}\cdot\epsilon^{\lambda'}_{\vecq}) +
    \frac{(\epsilon^{\lambda}_{\veck}\cdot\vecq)(\epsilon^{\lambda'}_{\vecq}\cdot\veck)}{q}\mC(|\vk+\vq|\eta')  \r\}\r|^2
\end{align}
Here, $\eta_k$ denotes the conformal time at which the wavenumber $k$ re-enters the horizon. The function $\mC(k,\eta)$ represents the transfer function governing the evolution of metric perturbations after inflation, while $\mC'(k,\eta)$ is its derivative with respect to conformal time. These are given by
\begin{align}
    \mC(k\eta_1) &= 2^{\beta+3/2} \Gamma\left[\beta + \frac{5}{2}\right] (k\eta_1)^{-\beta - 3/2} J_{\beta + 3/2}(k\eta_1), \\
    \mC'(k\eta_1) &= -\frac{1}{2}(k\eta_1) \Gamma\left[\beta + \frac{5}{2}\right]~\text{Hypergeometric0F1Regularized}\left[ \beta + \frac{7}{2}, -\frac{k^2 \eta_1^2}{4} \right].
\end{align}

Using this, we can express the number of photons produced per mode $k$, denoted by $N_k$, in terms of the metric power spectrum $P_{\Phi}(k)$ as
\begin{align}
    N_k = \frac{1}{2} \int_{\kpv}^{\ke} \frac{dq}{q} \left( \frac{q}{k} \right)^4 \frac{P_{\Phi}(|\vk + \vq|)}{|1 + \vq/\vk|^3} \mD(k, q),
\end{align}
where we have used the relation \( P_{\Phi}(k) = \frac{k^3}{2\pi^2} \Phi_0^2(k) \).

To proceed, we introduce a dimensionless variable \( u = q/k \), and define \( \gamma = \hat{k} \cdot \hat{q} = \cos(\theta) \), where \( \theta \) is the angle between the vectors \( \vk \) and \( \vq \).

In the above expression, the lower limit of the integral, \( \kpv \), corresponds to the pivot scale, associated with the longest observable wavelength today. The upper limit, \( \ke \), represents the highest wavenumber that can be excited by inflationary dynamics, i.e., the last mode that exited the horizon during inflation. Since we are primarily interested in observable (i.e., classical) perturbations, which become physically relevant only after horizon exit due to the quantum-to-classical transition, it is natural to choose \( \ke \) as the upper cutoff in the momentum integral.

We can rewrite the expression for $N_k$ in terms of the dimensionless variable $u = q/k$ and the cosine of the angle between $\vec{k}$ and $\vec{q}$, denoted as $\gamma = \hat{k} \cdot \hat{q} = \cos(\theta)$, as follows:
\begin{align}
    N_k = \int_{\umin}^{\umax} \frac{d u}{u} u^4 \int_{-1}^{1} d\gamma\, \frac{P_{\Phi}\left(k 
    (1 + u^2 - 2u\gamma)^{1/2}\right)}{(1 + u^2 - 2u\gamma)^{3/2}}\, \mD(k, u), \label{eq:def_nk_spec}
\end{align}
where for notational convenience, we have defined
\[
    \tilde{u} = \sqrt{1 + u^2 - 2u\gamma}.
\]

The function $\mD(k,q)$, which characterizes the gauge field amplification sourced by metric perturbations, can now be expressed in terms of $u$ and $\gamma$ as
\begin{align}
    \mD(k,q) = \left| \int_{\xe}^{x} dx \left\{ i\gamma \frac{\partial \mC(\tilde{u}, x)}{\partial x} + \left(1 + u\gamma - 2\gamma^2 \right) \mC(\tilde{u}, x) \right\} \right|^2.
\end{align}

In the above, we have used the relations
\[
    (\vec{\epsilon}^{\lambda}_{\vec{k}} \cdot \vec{\epsilon}^{\lambda'}_{\vec{q}}) = \cos(\theta) = \gamma, \qquad (\vec{\epsilon}^{\lambda}_{\vec{k}} \cdot \vec{q})(\vec{\epsilon}^{\lambda'}_{\vec{q}} \cdot \vec{k}) = \sin^2(\theta) = 1 - \gamma^2,
\]
and introduced the dimensionless integration variable \( x = k\eta' \). The transfer function $\mC(\tilde{u}, x)$ and its derivative govern the time evolution of the metric perturbation after inflation.

For convenience in evaluating the integral, we define two auxiliary integrals
\begin{align}
    \mI_1 = \int_{\xe}^{x} dx_1\, \frac{\partial \mC(\tilde{u}, x_1)}{\partial x_1}, \qquad 
    \mI_2 = \int_{\xe}^{x} dx_1\, \mC(\tilde{u}, x_1).
\end{align}

To evaluate the total spectrum \( N_k \), we divide the integral in Eq.~\eqref{eq:def_nk_spec} into two domains: \( u < 1 \) and \( u > 1 \), and rewrite the total contribution as
\begin{align}
    N_k &= \frac{1}{2} \int_{\umin}^{1} \frac{d u}{u} u^4 \int_{-1}^{1} d\gamma\, \frac{P_{\Phi}\left(k \sqrt{1 + u^2 - 2u\gamma}\right)}{(1 + u^2 - 2u\gamma)^{3/2}}\, \mD(k, u) \nonumber \\
    &\quad + \frac{1}{2} \int_{1}^{\umax} \frac{d u}{u} u^4 \int_{-1}^{1} d\gamma\, \frac{P_{\Phi}\left(k \sqrt{1 + u^2 - 2u\gamma}\right)}{(1 + u^2 - 2u\gamma)^{3/2}}\, \mD(k, u).
\end{align}

This decomposition is useful for numerical and analytical treatment, as the behavior of the integrand can vary significantly between the two domains.
\begin{align}
    N_k &=\int_{\umin}^1\frac{\d u}{u}u^4P_{\Phi}(k)\int_{-1}^1\d \gamma \,\mD_1(k,u)+\int_{1}^{\umax}\frac{\d u}{u}u^4\,\frac{P_{\Phi}(k\,u)}{u}\int_{-1}^1\d \gamma\,\mD_2(k,u)\\
    &=\int_{\umin}^1\frac{\d u}{u}u^4P_{\Phi}(k)\times\mF_1(k,u)+\int_{1}^{\umax}\frac{\d u}{u}u^4\,\frac{P_{\Phi}(k\,u)}{u^3}\times\mF_2(k,u)
\end{align}
Where $\mD_1(k,u)$ and $\mD_2(k,u)$ is define as
\begin{align}
    \mD_1(k,u)=\l|\int_{\xe}^{x}\d x_1\l(i\gamma\frac{\partial\mC(x_1)}{\partial x_1}+(1+u\gamma-2\gamma^2)\mC(x_1)\r\}\r|^2\\
    =\l| i\gamma\mI_{11}+(1+u\gamma-2\gamma^2)\mI_{12}\r|^2
\end{align}
\begin{align}
    \mD_2=\l| \int_{\xe}^{x}\d x_1 \l\{i\gamma \frac{\partial\mC(ux_1)}{\partial x} +(1+u\gamma-2\gamma^2)C(ux_1) \r\}\r|^2\\
    =\l| i\gamma\mI_{21}+(1+u\gamma-2\gamma^2)\mI_{22}\r|^2
\end{align}
where we defined
\begin{align}
    \mI_{11}=\int_{\xe}^{x}\d x_1\, \frac{\partial\mC(x_1)}{\partial x_1};\,\,\mI_{12}=\int_{\xe}^{x}\d x_1\, \mC(x_1);\,\, \mI_{21}=\int_{\xe}^{x}\d x_1\, \frac{\partial\mC(ux_1)}{\partial x_1};\,\, \mI_{22}=\int_{\xe}^{x}\d x_1\,\mC(ux_1)
\end{align}
Now it is easy to performed the $\gamma$ integral and we can write $F_1$ and $F_2$ as
\begin{align}
    F_1(k,u)=\frac{2}{3}\mI_{11}^2+\l(\frac{14}{15}+\frac{2u^2}{3}\r)\mI^2_{12}\\
    F_2(k,u)=\frac{2}{3}\mI^2_{21}+\l(\frac{14}{15}+\frac{2u^2}{3}\r)\mI_{22}^2
\end{align}
Now to perform the $\mI_{21}$ and $\mI_{22}$, we change the variable $\tilde{x}=ux$ and rewrite the integral as
\begin{align}
    \mI_{21}=\int_{\tx_i}^{\tx_0}\d \tx \frac{\partial\mC(\tx)}{\partial\tx};\,\,\, \mI_{22}=\frac{1}{u}\int_{\tx_i}^{\tx_0}\d \tx \mC(\tx)
\end{align}
\begin{figure}[t]
\begin{center}
\includegraphics[scale=0.4]{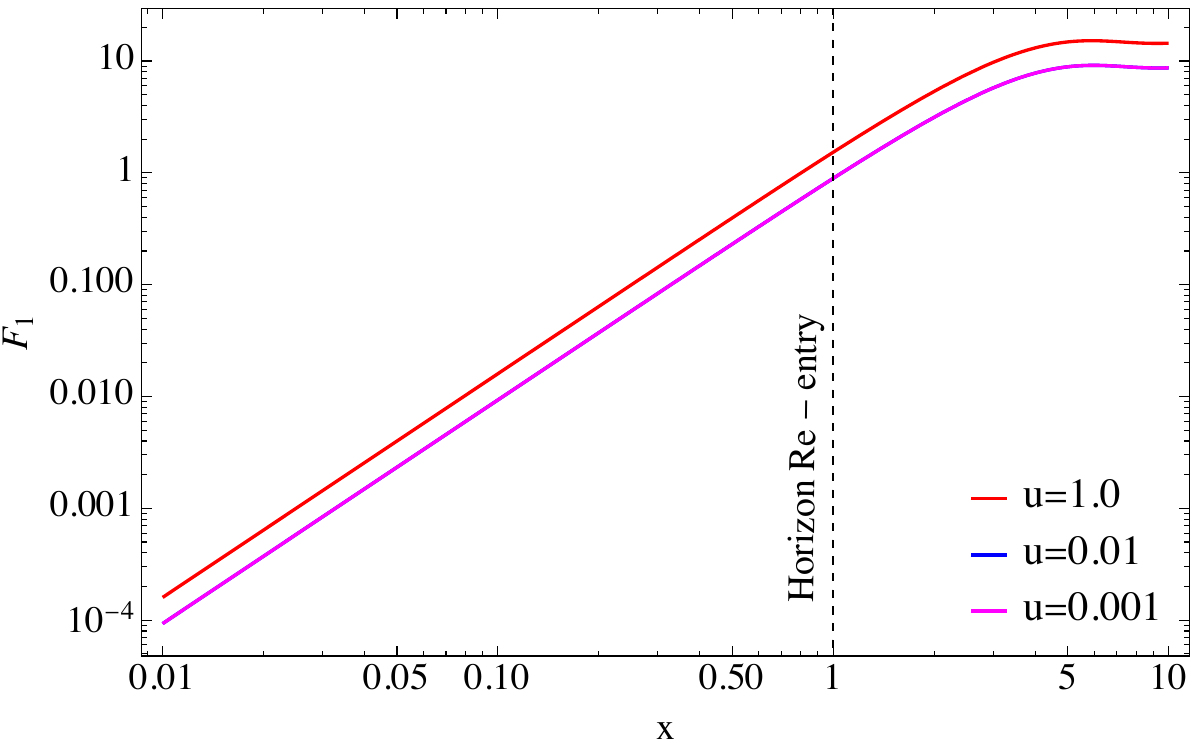}
\includegraphics[scale=0.4]{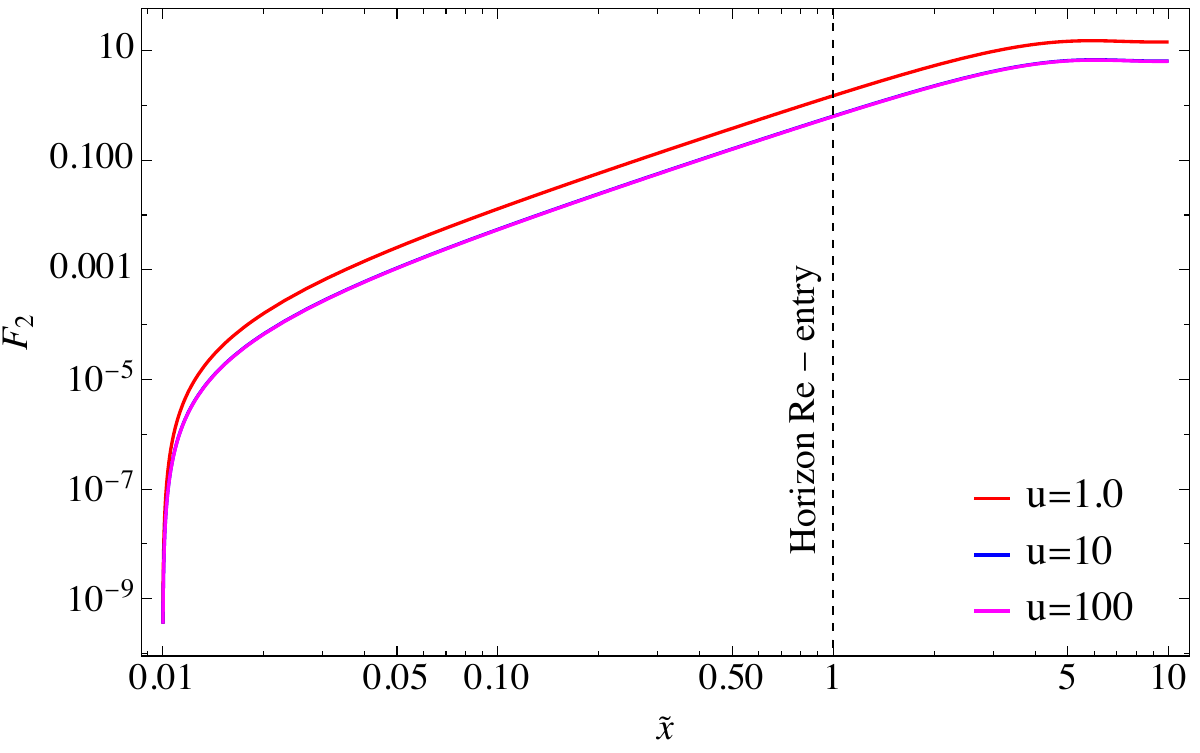}
 \caption{ 
 In the above figure, we present the time evolution of the functions \(F_1 \) and \( F_2 \), with three different colors representing different values of the ratio \( u = q/k \). The vertical dashed black line in both panels marks the moment of horizon re-entry for the corresponding mode.
    }
    \label{fig:F}
\end{center}
\end{figure}
Since the integrals \( \mI_{21} \) and \( \mI_{22} \) are computed in the regime where \( u \geq 1 \), corresponding to \( q > k \), we can safely extend the upper limit of the integration to \( x \gtrsim 1 \). Numerical evaluations confirm that both \( \mathcal{I}_1 \) and \( \mathcal{I}_2 \) rapidly saturate for \( x \gtrsim 1 \), indicating that gauge field production effectively ceases once the mode re-enters the horizon. This behavior implies that the amplification of the gauge field is primarily driven by super-horizon metric perturbations. After horizon re-entry, the efficiency of this process diminishes significantly due to damping effects, such as the large conductivity of the post-inflationary plasma.

After inflation, the metric perturbation spectrum \( P_{\Phi}(k) \) and the curvature power spectrum \( \mathcal{P}_{\mathcal{R}}(k) \) are related via Eq.~\eqref{eq:P_phi_post_i}. Using this relation, we can express the number spectrum \( N_k \) directly in terms of the curvature perturbation spectrum as
\begin{align}
    N_k=\l(\frac{2+\beta}{3+2\beta}\r)^2\l[ \int_{\umin}^1\frac{\d u}{u}u^4\mP_{\mR}(k)\times F_1(k)+\int_{1}^{\umax}\d u\,\mP_{\mR}(k\,u)\times F_2(k)\r]\label{eq:Nk_post_inf}
\end{align}
\subsection{Number spectrum due to Inflationary curvature power spectrum:}
Let us now consider a simple slow-roll inflationary scenario, where the curvature power spectrum generated by inflaton fluctuations is given by
\begin{align}
    \mathcal{P}_{\mathcal{R}}(k) = A_s \left( \frac{k}{\kpv} \right)^{n_s - 1}, \label{eq:PR_powerlaw}
\end{align}
with \( A_s \) denoting the amplitude of the primordial scalar power spectrum and \( n_s \) representing the scalar spectral index.

Substituting Eq.~\eqref{eq:PR_powerlaw} into the expression for the number spectrum \( N_k \) derived earlier in Eq.~\eqref{eq:Nk_post_inf}, we obtain
\begin{align}
    N_k &=\frac{\mF_1(k)}{8}\mP_{\mR}(k)\l\{ 1-\l(\frac{\kpv}{k}\r)^4\r\}+\frac{\mF_2}{2}A_s\l(\frac{k}{\kpv}\r)^{\ns-1}\int_1^{\umax}\d u u^{\ns-1}\\
    &=\frac{\mF_1(k)}{8}\mP_{\mR}(k)\l\{ 1-\l(\frac{\kpv}{k}\r)^4\r\}+\frac{\mF_2}{2}A_s\l(\frac{k}{\kpv}\r)^{\ns-1}\frac{1}{\ns}\l\{ \l(\frac{\ke}{k}\r)^{\ns}-1\r\}\label{eq:nk_inflationary}
\end{align}
Observational data suggest that the scalar spectral index takes the value \( n_s \simeq 0.975 \)~\cite{Planck:2018vyg}. Substituting this into Eq.~\eqref{eq:nk_inflationary}, we can approximate the number spectrum as
\begin{align}\label{eq:nk_v}
    N_k \simeq \left( \frac{2 + \beta}{3 + 2\beta} \right)^2 \left\{ 
    \frac{\mathcal{F}_1(k)}{8} A_s \left( \frac{k}{\kpv} \right)^{n_s - 1} 
    + \frac{\mathcal{F}_2}{2n_s} A_s \left( \frac{\ke}{\kpv} \right)^{n_s} \left( \frac{\kpv}{k} \right) \right\},
\end{align}
where \( \mathcal{F}_1(k) \) and \( \mathcal{F}_2 \) are kernel-dependent integrals defined previously, and \( k_{\mathrm{end}} \) corresponds to the highest wavenumber that exited the horizon during inflation. Since \( n_s \lesssim 1 \), the second term in Eq.~\eqref{eq:nk_v} dominates over the first, leading to the effective scaling \( N_k \propto k^{-1} \) for the number spectrum in this scenario.

\medskip

In contrast, ultra-slow-roll inflation models typically produce an enhanced curvature power spectrum. For illustrative purposes, we consider the following parameterized form
\begin{align}
    \mathcal{P}_{\mathcal{R}}(k) = A_0 \exp\left[ -\frac{(k - k_p)^2}{\delta \, k_p^2} \right],
\end{align}
where \( A_0 \) is the peak amplitude, \( k_p \) is the peak scale, and \( \delta \) controls the width of the enhancement. Substituting this into the general expression for \( N_k \), we obtain
\begin{align}\label{eq:nk_s}
    N_k = \left( \frac{2 + \beta}{3 + 2\beta} \right)^2 \left\{ 
    \frac{\mathcal{F}_1(k)}{8} \mathcal{P}_{\mathcal{R}}(k)
    + \sqrt{\pi \delta} A_0 \mathcal{F}_2(k) \left( \frac{k_p}{k} \right) 
    \left[ 1 + \mathrm{Erf}\left( \frac{k_p - k}{k_p \sqrt{\delta}} \right) \right] \right\}.
\end{align}

\medskip

From Eq.~\eqref{eq:nk_s}, it is evident that for \( k < k_p \), the dominant contribution to \( N_k \) again scales as \( N_k \propto k^{-1} \), consistent with the slow-roll case. Therefore, regardless of the detailed shape of the primordial curvature spectrum, whether slow-roll type or PBH-type, the number of photons produced per comoving mode exhibits an inverse scaling with \( k \).

\medskip

Next, we compute the magnetic energy density associated with a given comoving mode \( k \), evaluated at the present epoch.
\begin{align}
    \rho_{\rm B}(k,\eta_0)=\l(\frac{k}{a(\eta_0)}\r)^4N_k(\eta)\propto k^3
\end{align}
Now the present-day magnetic field strength is defined as $B_0(k)\propto \sqrt{\rhob(k)}\propto k^{3/2}$.

We now aim to determine the threshold value of the peak wavenumber \( \kp \) such that the gauge field production is dominated by the ultra-slow-roll (PBH-type) curvature power spectrum, rather than the simple slow-roll type curvature power spectrum. To establish this criterion, we compare the number spectra given in Eqs.~\eqref{eq:nk_v} and \eqref{eq:nk_s}. Equating the dominant contributions from both expressions yields the condition
\begin{align}
    \kp^{\rm c}\simeq\frac{A_s\ke}{2\sqrt{\pi\delta}\ns A_0}\l(\frac{\ke}{\kpv}\r)^{\ns-1}.
\end{align}

\bibliographystyle{apsrev4-1}
\bibliography{references}

\end{document}